\documentclass[fleqn,usenatbib]{mnras}

\usepackage{newtxtext,newtxmath}
\usepackage[T1]{fontenc}
\usepackage{graphicx}	
\usepackage{amsmath, bm}
\usepackage{mydefs}
\usepackage{multirow}
\usepackage{booktabs}
\usepackage{chemformula}
\usepackage[modulo]{lineno}
\usepackage{orcidlink}
\usepackage{ulem}

\DeclareRobustCommand{\VAN}[3]{#2}
\let\VANthebibliography\thebibliography
\def\thebibliography{\DeclareRobustCommand{\VAN}[3]{##3}\VANthebibliography}

\title[CSST BAO with slitless spec-z errors]{The effect of slitless spectroscopic redshift uncertainty and interlopers on BAO measurements from CSST-like samples}

\author[Zhejie Ding]{
Zhejie Ding,\orcidlink{0000-0002-3369-3718}$^{1,2}$\thanks{zjding@niaot.ac.cn}
Yu Yu,\orcidlink{0000-0002-9359-7170}$^{3,4}$\thanks{yuyu22@sjtu.edu.cn}
Pengjie Zhang $^{3,4,5}$
\\
$^{1}$Nanjing Institute of Astronomical Optics \& Technology, Chinese Academy of Sciences, Nanjing 210042, China\\
$^{2}$University of Chinese Academy of Sciences, Nanjing 211135, China\\
$^{3}$Department of Astronomy, School of Physics and Astronomy, Shanghai Jiao Tong University, Shanghai 200240, China\\
$^{4}$Key Laboratory for Particle Astrophysics and Cosmology (MOE)/Shanghai Key Laboratory for Particle Physics and Cosmology, Shanghai 200240, China\\
$^{5}$Tsung-Dao Lee Institute, Shanghai Jiao Tong University, Shanghai 201210, China\\
}

\date{Accepted XXX. Received YYY; in original form ZZZ}

\pubyear{\the\year{}}

\begin{document}
\label{firstpage}
\pagerange{\pageref{firstpage}--\pageref{lastpage}}
\maketitle

\begin{abstract}
The baryon acoustic oscillations (BAO) is a vital probe to measure cosmological distances and constrain dark energy. The ongoing and near-future space-based telescopes, including the Chinese Space Station Survey Telescope (CSST), will perform galaxy redshift surveys using slitless spectroscopy. 
Due to the relatively low spectral resolution and observational effects, there will be uncertainties and interloper contamination in measured redshifts. In this study, we consider the Gaussian-distributed redshift uncertainty and [O III]-H$\beta$ interlopers in the CSST-like slitless spectroscopic redshift (spec-z) samples, which are simulated using the approximate $N$-body code \textsc{FastPM}. 
We study the effect of spec-z errors on BAO before and after density field reconstruction.
Both the redshift uncertainty and interlopers can damp the BAO signal, and the latter can further induce oscillations in the broadband shape of the power spectrum. We model the interloper effect on the propagator, i.e. the cross-correlation between the observed and initial density fields, as well as on the BAO power spectrum. We study the influence of the spec-z errors on the fitted anisotropic BAO parameters $\alperp$ and $\alpara$. The resulting systematic bias on $\alperp$ and $\alpara$ is mild, mostly within $0.1$ per cent and $0.2$ per cent, respectively. Even with the spec-z errors, BAO reconstruction can still significantly reduce the systematic bias and statistical errors on the $\alpha$ parameters. Furthermore, we vary the redshift uncertainty and study the reconstruction performance. 
\end{abstract}


\section{Introduction}
The baryon acoustic oscillations (BAO) originated from the interaction between photons and baryons in the early Universe, and the relics remain in the matter distribution in the late Universe \citep{Peebles_Yu_1970,Sunyaev_Zeldovich_1970}. Since the first BAO detection from galaxy clustering two decades ago \citep{Cole2005, Eisenstein2005}, BAO has become one of the most important probes to measure cosmological distances and constrain dark energy \citep{Weinberg2013}. It has been routinely measured in multiple galaxy surveys across wide redshift ranges and increasing statistical constraining power \citep[e.g.][]{Beutler2011, Blake2011,Alam2017,Alam2021}. The recent BAO measurements from the Dark Energy Spectroscopic Instrument (DESI), combined with the cosmic microwave background or Type Ia supernovae, result in a significant signature of dynamical dark energy \citep{DESI_DR1_BAO,DESI_DR1_Lya,DESI_DR1_cosmo,2025arXiv250314738D}. To validate the DESI finding, it is significant to obtain BAO measurements from other galaxy redshift surveys, such as the Subaru Prime Focus Spectrograph \citep{Takada2014} and the 4-metre Multi-Object Spectroscopic Telescope \citep{deJong2019}, as well as space-based telescopes, including Euclid \citep{Euclid2025}, the Nancy Grace Roman Space Telescope \citep[hereafter Roman,][]{Spergel2015}, and the Chinese Space Station Survey Telescope \citep[CSST,][]{Zhan2011,Zhan2021,Gong2025b}. 

Euclid, Roman, and CSST will all perform slitless spectroscopic redshift (hereafter spec-z) surveys. For the case of CSST, it covers the wavelength range $255 - 1000$ nm, split into three grating bands, i.e. $GU$, $GV$, and $GI$ \citep{Gong2025b}. The average spectral resolution is $R\gtrsim 200$. The CSST spec-z survey covers the redshift range $0.0<z<1.0$. For slitless spectroscopy, detecting emission lines is critical to calibrate target redshifts. Due to the relatively low spectral resolution, emission lines are broadened, introducing redshift uncertainty in spectral fitting. For CSST, the spec-z uncertainty is estimated to be $\gtrsim 0.002(1+z)$ \citep{Gong2019,Zhang2026}, which is about an order of magnitude larger than that of fiber-based spectroscopic surveys, such as DESI \citep{2022AJ....164..207D}.

In addition, line interlopers are common in slitless spectroscopy. \cite{Wen2024} simulated CSST spectra and found that a large number of galaxy spectra contain only one emission line with a reasonable signal-to-noise (S/N) ratio due to various instrumental and observational effects. 
\cite{2025RAA....25b5015C} furthermore showed that morphological broadening can cause nearby emission lines to overlap. All these effects can make one emission line misidentified as another, inducing incorrect redshift measurements. The misidentified tracers are referred as interlopers, which introduce systematics in the galaxy clustering of interest \citep[e.g.][]{Pullen2016,Addison2019,He2025,Risso2026}. There have been several studies modelling or reconstructing the interloper fraction to mitigate their impact \citep[e.g.][]{Grasshorn_Gebhardt2019,Farrow2021, Gong2021, Foroozan2022,Peng2023}.

In the CSST spec-z range, there are several prominent emission lines, including [O II], H$\gamma$, $\Hbeta$, \oiii{}, and H$\alpha$ \citep{Sui2025}. Among different types of redshift interlopers that may occur in CSST, we specifically focus on the \oiii{}-$\Hbeta$ interlopers, i.e. $\Hbeta$ $\lambda 4861$ misidentified as \oiii{} $\lambda 5007$.\footnote{\oiii{} $\lambda\lambda 4959,5007$ have two emission lines. Only the latter is considered because of the stronger signal. \citet{Foroozan2022} referred to the \oiii{}-$\Hbeta$ interlopers as small-displacement interlopers, since the comoving displacement of a $\Hbeta$ interloper is $<150\Mpch$, and defined other types of interlopers with a larger wavelength difference between target and interloper lines as large-displacement interlopers. In this work, we adopt their notation, using the \oiii{}-$\Hbeta$ and small-displacement interlopers interchangeably.} Due to the close wavelengths of \oiii{} and $\Hbeta$, the redshift offset of a misidentified $\Hbeta$ galaxy from its true redshift is relatively small; hence, there is cross-correlation between the \oiii{} targets and $\Hbeta$ interlopers in a given redshift bin. 
It turns out that the \oiii{}-$\Hbeta$ displacement would be close to the BAO scale ($\sim100~\Mpch$), inducing a peak in the galaxy correlation function near the BAO peak. Without modelling the effect, it can significantly contaminate the BAO measurement \citep{Massara2021,Foroozan2022,Nguyen2024}.
Unlike previous work, we focus on the BAO signal only, assuming that the broadband shape of the galaxy power spectrum is well modelled. After including the interloper effect in the fitting model, we study whether such interlopers can bias the BAO measurement at high precision.

As a caveat, we ignore other types of redshift interlopers, including noise spikes and large-displacement interlopers. The former likely merely add shot noise and change only the amplitude of the target power spectrum. The latter have a relatively large redshift offset from the targets; hence, the cross-correlation between targets and interlopers is small. If the redshift and fraction of interlopers are known, the interloper clustering can be modelled and removed from the contaminated clustering, as discussed, e.g. by \citet{Foroozan2022,Risso2026}.

Density field reconstruction (also called BAO reconstruction) has become a standard tool widely applied to BAO measurements from real data \citep[e.g.][]{Wang2017, Bautista2021, DESI_DR1_BAO}. BAO reconstruction was first proposed by \cite{Eisenstein2007b}, based on the Zel'dovich approximation \citep{Zeldovich1970}. It helps to largely restore BAO signal smeared by nonlinear structure growth and redshift space distortions (RSD), and to reduce the systematics in BAO scale parameters. BAO reconstruction has been widely studied both theoretically and in simulations \citep[e.g.][]{Seo2008,Padmanabhan2009,Noh2009,Burden2014,White2015,Seo2016,CDP2024,Paillas2024}.
We study the influence of CSST slitless spec-z errors on BAO reconstruction. Recently, \citet{Shi2025} forecasts CSST BAO measurement before reconstruction, using $N$-body simulation with the added redshift uncertainty. Our study can provide insights into the reconstruction efficiency, which is useful for more reliable BAO forecasts from theoretical modelling \citep{Ding2024, Miao2024}.

This paper is structured as follows. In Section \ref{sec:method}, we present the construction of the CSST-like mock samples with slitless spec-z errors using \textsc{FastPM} simulations. We describe the methodology of BAO reconstruction, propagator, and BAO fitting model. In Section \ref{sec:results}, we show our main results, including the BAO power spectrum, nonlinear damping parameters, and anisotropic BAO scale parameters fitted from the simulations. In Section \ref{sec:conclusion}, we present our conclusions and discussions.

\section{Methodology}
\label{sec:method}

\subsection{Mock samples}
\label{sec:sim}
We use \textsc{FastPM} \citep{Feng2016} simulations to construct dark matter halo mocks, which mimic \textit{CSST}-like spectroscopic galaxy samples in terms of galaxy number density and bias. \textsc{FastPM} is an approximate $N$-body code, using a particle-mesh $N$-body solver with modified kick and drift factors to enforce correct linear displacement evolution. With a small number of time steps and coarse force resolution, \textsc{FastPM} is able to recover the matter power spectrum at the percent level up to $k=1.0\hMpc$, compared to that from a regular $N$-body simulation.  
We use the \textsc{FastPM} simulations created in our previous work \citep{Ding2018}. These simulations adopt the Planck 2015 flat $\Lambda$CDM cosmology \citep{Planck2015}, with $\Omega_\text{m}=0.3075$, $\Omega_{\text{b}}=0.0486$, $h=0.6774$, and $\sigma_8=0.8159$. The simulation box has a side length of $1380$ $\Mpch$, and contains $2048^3$ dark matter particles with a mass resolution of $2.6\times 10^{10}\Msunh$. 
The simulation outputs the dark matter density field at snapshots $z=2.5$, $2.0$, $1.5$, $1.0$, $0.6$ and $0$.
The halo catalogues are constructed with the friends-of-friends halo finder, which is implemented in \textsc{nbodykit}\footnote{\url{https://nbodykit.readthedocs.io/en/latest/index.html\#}} \citep{Hand2018}. We set the linking length equal to $0.2$ times the mean separation of dark matter particles.

We apply sample variance cancellation to the paired simulations to increase the statistical constraining power \citep{Schmittfull2015, Prada2016}. Each pair of simulations shares the same white noise in its initial conditions (IC): one simulation uses the input linear power spectrum containing the BAO signal, while the other uses the power spectrum with the BAO signal smoothed away. 
Taking the difference of the power spectra from each pair cancels most of the sample variance induced by broadband fluctuations; hence, it significantly increases the S/N of the BAO measurement. In this study, we use 90 pairs of simulations.

Based on the expected galaxy number density from the CSST slitless spec-z survey \citep{Gong2019}, we select the most massive haloes to match the number density at redshifts $z=0$, $0.6$ and $1.0$. These snapshots are representative of the CSST slitless spec-z range, and the galaxy number density decreases from low to high redshifts. At a given redshift, we select massive haloes with a specific mass cut, and show the number density of the selected haloes in Table~\ref{tab:halo_sample}. We compute the halo bias from the halo power spectra at large scales. As a result, the halo bias is close to the expected galaxy bias, for which we adopt the simple formula $b_g = 1 + 0.84z$ \citep{Weinberg2004}. We note that \citet{Pei2024} forecasts \oiii{} galaxy bias as a function of luminosity and redshift, which is more realistic.
We have not considered any selection of galaxies with high S/N spectra or emission lines, which can reduce our assumed galaxy number density \citep{Wen2024, Sui2025, Peng2026}. We leave a more detailed study for future work.

\begin{table}
	\centering
	\caption{The \textsc{FastPM} dark matter halo mocks are used to construct CSST-like galaxy samples from the slitless spec-z survey. We select three redshift snapshots representative of the survey redshift coverage. At each redshift, we show the halo mass cut, number density, bias, linear growth rate, and interloper displacement.}
	\label{tab:halo_sample}
	\begin{tabular}{|c|c|c|c|c|c|} 
		\hline
		z & $M_{\text{cut}}$ [$\Msunh$] & $n_\text{halo}$ [$\hMpccube$] & bias & f & $\Delta d$ [$\Mpch$] \\
        \hline
        0.0 & $8.6\times 10^{11}$ & $4.5\times 10^{-3}$ & 0.84 & 0.52 & 89.4\\
		0.6 & $1.9\times 10^{12}$ & $2.1\times 10^{-3}$ & 1.36  & 0.79 & 101.5\\
        1.0 & $5.85\times 10^{12}$ & $5.0\times 10^{-4}$ & 2.25   & 0.87 & 99.5\\
		\hline
	\end{tabular}
\end{table}

\subsection{Slitless spec-z errors}
We consider two sources of CSST slitless spec-z errors in this study. One is redshift uncertainty, which arises from the relatively low resolution of slitless spectroscopy. We assume a Gaussian distribution for the uncertainty, i.e. $\mathcal{N}(0,\, \sigma^2_{\text{spec-z}})$, where $\sigma_{\text{spec-z}}=\sigma_0(1+z)$. For the default case, we set $\sigma_0=0.002$. 
The other source is the \oiii{}-$\Hbeta$ small-displacement interlopers. The interloper displacement is given by
\begin{align}
    d^{\text{int}} &= \int_{z_1}^{z_2}\frac{cdz}{H(z)}, \label{eq:displacement_interloper}
\end{align}
where $H(z)$ is the Hubble parameter, and $z_1$ and $z_2$ are the true redshifts of the \oiii{} ($\lambda_1=500.7$ nm) and $\Hbeta$ ($\lambda_2=486.1$ nm) targets, respectively.

In addition to the spec-z errors considered in this work, the observed redshift is always affected by the intrinsic peculiar velocities of galaxies, which give rise to RSD in galaxy clustering \citep{Kaiser1987}. Since we use cubic-box simulations, we fix the line of sight (LoS) along the $z$-axis. Accounting for all these effects, we simulate the observed position along the LoS from the true position as
\begin{align}
    r^{\text{LoS}}_{\text{obs}} = r^{\text{LoS}}_{\text{true}} + d^{\text{RSD}} + d^{\text{Gzerr}} + d^{\text{int}},\label{eq:distance_LoS}
\end{align}
where $r^{\text{LoS}}_{\text{true}}$ is the true position along the LoS in real space. $d^{\text{RSD}}$ accounts for RSD, i.e.
\begin{align}
d^{\text{RSD}}=\frac{v_z(1+z)}{H(z)},    
\end{align}
where $v_z$ is the velocity component along the LoS. $d^{\text{Gzerr}}$ is the contribution from the Gaussian spec-z uncertainty, i.e.
\begin{align}
d^{\text{Gzerr}}=\frac{c\sigma_z}{H(z)}. \label{eq:displacement_Gzerr} 
\end{align}
At $z=0.0$, $0.6$, and $1.0$, the comoving-distance uncertainty $d^{\text{Gzerr}}$ is $6.0\Mpch$, $6.9\Mpch$, and $6.8\Mpch$, respectively.

To study interloper systematics, we consider three values of the interloper fraction $I_\text{frac}$, i.e. $1\%$, $5\%$ and $10\%$. We randomly select a fraction $I_\text{frac}$ of the targets as interlopers from the parent mocks. During this selection, we set the same initial seed of the random number generator for a given pair of mocks. We apply periodic boundary conditions to the positions after applying the displacements in equation (\ref{eq:distance_LoS}). In the following analysis, whether or not we include $d^{\text{Gzerr}}$ and $d^{\text{int}}$, we always consider RSD and take the RSD-only case as the default. Comparing the results to the default one, we can quantify the systematic effects on the BAO measurement due to spec-z errors.

\subsection{Galaxy power spectrum with small-displacement interlopers}
Following \citet{Pullen2016} and \citet{Foroozan2022}, we derive the galaxy power spectrum with small-displacement interlopers. For the galaxy field with interlopers, the observed overdensity is
\begin{align}
    \delta_\text{obs}(\bm{r})=(1-I_\text{frac})\delta_\text{t}(\bm{r}) + I_\text{frac}\delta_\text{int}(\bm{r}),\label{eq:delta_obs}
\end{align}
where $\delta_\text{t}$ and $\delta_\text{int}$ are the target and interloper overdensities, respectively. With the line of sight fixed, the observed position of an interloper is
\begin{align}
    \bm{r} = \bm{r'} + d^{\text{int}}\hat{z},
\end{align}
where $\bm{r'}$ is the true position of the interloper. In a given redshift bin, the small displacement $d^{\text{int}}$ can be approximated as a constant. In addition, assuming that the galaxy bias of interlopers is the same as that of the target sample, we have $\delta_\text{int}(\bm{r})=\delta_\text{t}(\bm{r'})$. In Fourier space,
\begin{align}
    \delta_\text{int}(\bm{k}) &= \int \delta_\text{int}(\bm{r})\exp(-i \bm{k}\bm{r}) d^3r \nonumber \\
    &= \int \delta_\text{t}(\bm{r'})\exp[-i\bm{k}(\bm{r'} + d^{\text{int}}\hat{z})] d^3r' \nonumber \\
    &= \delta_\text{t}(\bm{k})\exp(-ikd^\text{int}\mu),\label{eq:deltak_int}
\end{align}
where $\mu$ is the cosine of the angle between $\bm{k}$ and the line of sight.
The observed galaxy power spectrum is given by
\begin{align}
    &P_\text{obs}(\bm{k})(2\pi)^3\delta_\text{3D}(\bm{k} + \bm{k'}) = \langle \delta_\text{obs}(\bm{k})\delta_\text{obs}(\bm{k'})\rangle \nonumber \\
    &= (1-I_\text{frac})^2\langle \delta_\text{t}(\bm{k})\delta_\text{t}(\bm{k'}) \rangle + (1-I_\text{frac})I_\text{frac} [\langle\delta_\text{t}(\bm{k})\delta_\text{int}(\bm{k'}) \rangle + \nonumber \\ &\;\;\;\;\; \langle\delta_\text{t}(\bm{k'})\delta_\text{int}(\bm{k}) \rangle ] + I^2_\text{frac} \langle \delta_\text{int}(\bm{k})\delta_\text{int}(\bm{k'}) \rangle,\label{eq:Pkobs}
\end{align}
where $\delta_\text{3D}$ is the 3D Dirac delta function, which enforces $\bm{k}=-\bm{k'}$. Substituting equation (\ref{eq:deltak_int}) into equation (\ref{eq:Pkobs}), we obtain
\begin{align}
    P_\text{obs}(k, \mu) = \left[1 - 2I_\text{frac}(1- I_\text{frac})\big(1 - \text{cos}(kd^\text{int}\mu)\big)\right] P_\text{tt}(k, \mu), \label{eq:Pobs_int}
\end{align}
where $P_\text{tt}$ is the target galaxy power spectrum in the absence of interlopers. The power spectrum multipoles can be calculated via
\begin{align}
    P_{\ell}(k) = \frac{2\ell+1}{2}\int_{-1}^{1}P(k, \mu) L_{\ell}(\mu) d\mu,
\end{align}
where $L_{\ell}(\mu)$ is the Legendre polynomial of degree $\ell$. 
We show the power spectrum multipoles with spec-z errors in Fig. \ref{fig:pkmu_now_z0.6}, and compare those with small-displacement interlopers to the model.

\subsection{BAO reconstruction}
Due to nonlinear structure growth, the Eulerian position $\bm{r}$ of a galaxy is displaced from the initial position $\bm{q}$ by a displacement field $\bm{\Psi}$, given by
\begin{align}
\bm{r}=\bm{q}+\bm{\Psi}.   
\end{align}
The RSD effect further smears the observed position along the LoS. Together, they can shift and degrade the observed BAO peak from its true position; if not corrected, these effects can introduce systematics in the constraints on cosmological distances.

Based on the Zel'dovich approximation, the standard BAO reconstruction estimates the real-space displacement field from the observed nonlinear galaxy number density field by solving the continuity equation, i.e.  
\begin{align}
    \nabla \cdot \bm{\Psi} + \frac{f}{b} \nabla\cdot (\bm{\Psi}\cdot \bm{\hat{r}})\bm{\hat{r}} = -\frac{\delta^{\text{s}}_{\text{obs}}(\bm{r})}{b}, \label{eq:Zeldovich_app}
\end{align}
where $f$ is the linear growth rate $\frac{d\ln D}{d\ln a}$ with $D$ as the linear growth factor and $a$ as the scale factor, $b$ is the linear galaxy bias, and $\bm{\hat{r}}$ is the LoS unit vector. $\delta^{\text{s}}_\text{obs}$ denotes the observed galaxy density contrast with the addition of a smoothing kernel. In Fourier space, 
\begin{align}
    \delta^{\text{s}}_{\text{obs}}(\bm{k}) = \delta_{\text{obs}}(\bm{k})S(k),
\end{align}
where $S(k)$ is the smoothing kernel, which is traditionally set as Gaussian, i.e.
\begin{align}
    S(k)= \exp\left(-k^2\Sigma^2_{\text{sm}}/2\right).
\end{align}
We set the density smoothing scale $\Sigma_{\text{sm}}=15\Mpch$, based on \cite{CDP2024} and \cite{Paillas2024}, who conducted extensive studies on parameter settings in the standard reconstruction scheme. 

For a fixed LoS, $\bm{\Psi}$ can be solved straightforwardly from equation (\ref{eq:Zeldovich_app}), i.e.
\begin{align}
    \bm{\Psi}=-\frac{i\bm{k}\delta^{\text{s}}_{\text{obs}}(\bm{k})}{b (1+\beta \mu^2)k^2},
\end{align}
where $\beta=f/b$. To obtain the reconstructed density field, we displace the mock galaxies by $-(1+f)\bm{\Psi}$ and get the displaced density field, denoted as $\delta_{\text{disp}}(\bm{x})$. To reconstruct the large-scale information correctly, we sample a set of random particles and shift them by the same amount as the galaxies. Considering computational resources and data storage, we set the random sample size 10 times the data size. To check whether the random sample size can bias the BAO measurement, we have compared the result using a random sample 20 times the data size and found little difference. The random-shifting convention we adopt is denoted as RecSym, which retains RSD in the reconstructed field. Relative to the other convention denoted as RecIso, which shifts the random particles by $-\bm{\Psi}$, the RecSym convention appears less sensitive to errors in the reconstructed displacement \citep{CDP2024}. With the shifted random denoted as $\delta_{\text{shift}}(\bm{x})$, the reconstructed density field is defined as $\delta_{\text{rec}} \equiv \delta_{\text{disp}} - \delta_{\text{shift}}$. For both cases with and without the slitless spec-z errors, we perform the BAO reconstruction for each simulation using \textsc{pyrecon}\footnote{\url{https://github.com/cosmodesi/pyrecon}}. We study whether there is any systematics in the reconstructed BAO signal induced by the slitless spec-z errors.

\subsection{Propagator}
The propagator measures the cross-correlation between the initial linear density field $\delta_{\mathrm{lin}}$ and the final (observed) density field $\delta_{\mathrm{obs}}$ in Fourier space \citep[e.g.][]{Crocce2006,Crocce2008}. In this study, we use the propagator to assess the BAO reconstruction efficiency \citep[e.g.][]{Seo2010}. At a given redshift $z$, we can define the normalized propagator as
\begin{align}
    G(k, \mu) = \frac{1}{b (1+\beta \mu^2)} \frac{\langle \delta_{\mathrm{lin}}(\bm{k}, z) \delta^*_{\mathrm{obs}}(\bm{k}, z) \rangle}{\langle\delta_{\mathrm{lin}}(\bm{k}, z)  \delta^*_{\mathrm{lin}}(\bm{k}, z) \rangle}.\label{eq:propagator}
\end{align}
Without interlopers, the Gaussian damping model fits the observed propagator well, i.e.
\begin{align}
G_{\text{model}}(k, \mu) = \exp\left[-\frac{1}{4}k^2\Big((1-\mu^2)\Sigma^2_{\perp} + \mu^2\Sigma_{\|}^2\Big)\right], \label{eq:model_prop}
\end{align}
where $\Sigma_{\perp}$ and $\Sigma_{\|}$ are the nonlinear BAO damping parameters perpendicular and parallel to the LoS, respectively. Measuring the propagator is one method to estimate $\Sigmaperp$ and $\Sigmapara$ from simulations. The prior ranges of $\Sigmaperp$ and $\Sigmapara$ can then be set in the BAO fit for real data analysis \citep[e.g.][]{DESI_DR1_BAO}.

With small-displacement interlopers in $\delta_{\mathrm{obs}}$, the propagator model can be derived based on equations (\ref{eq:delta_obs}) and (\ref{eq:deltak_int}), i.e. 
\begin{align}
    G^{\text{int}}_{\text{model}}(k, \mu) = \left[1-I_{\text{frac}}\big(1-\cos(kd^\text{int}\mu)\big)\right] G_{\text{model}}(k, \mu). \label{eq:model_prop_int}
\end{align}
For the case with BAO reconstruction, we expect that the interloper impact on the density field to be smaller than in the pre-reconstruction case; however, the exact modelling is left for future study. For now, we naively adopt the same model.

\begin{figure*}
    \includegraphics[width=1.95\columnwidth]{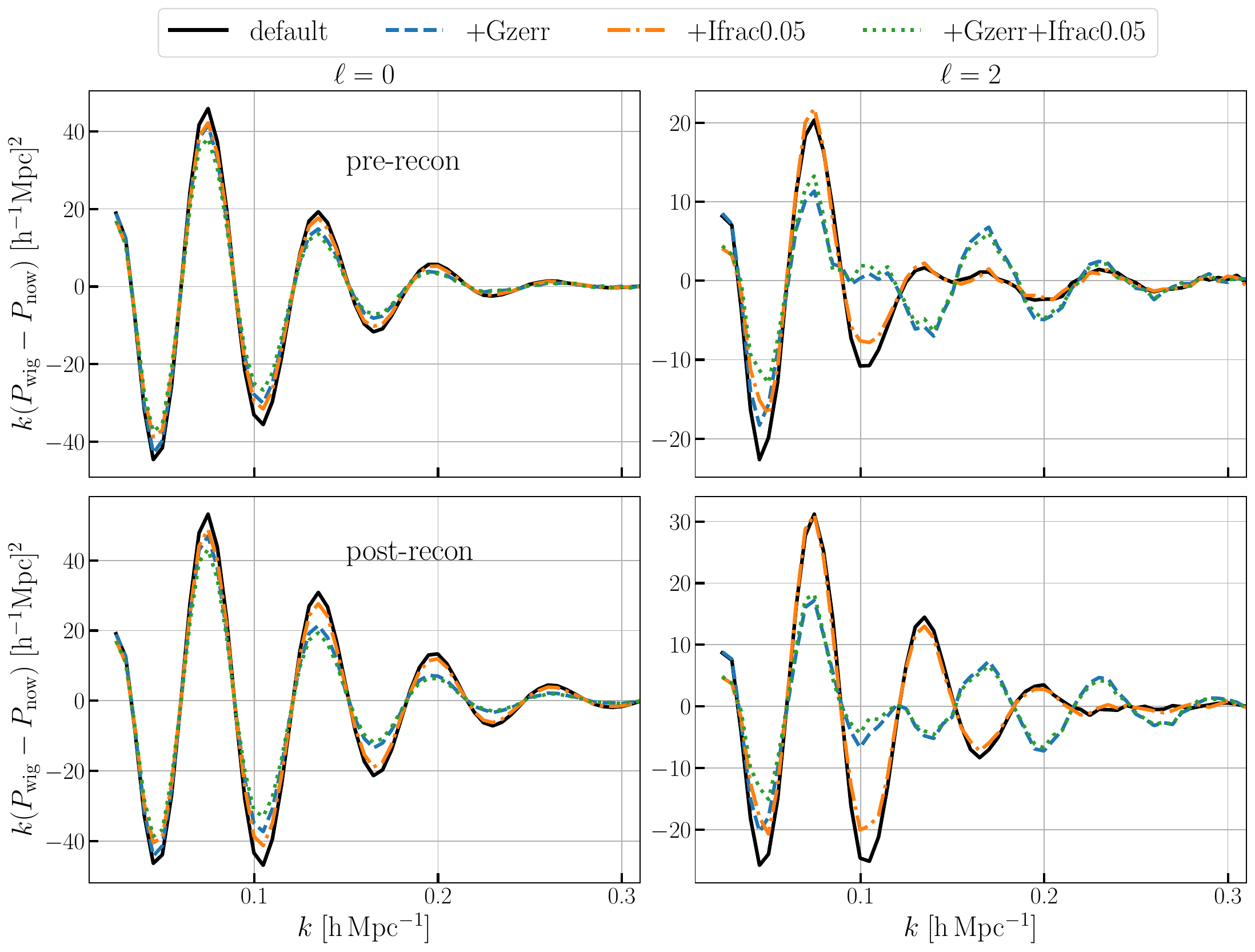}
    \caption{Effect of slitless spec-z uncertainty and interlopers on the BAO power spectra at $z=0.6$. We study the cases with Gaussian redshift uncertainty $\sigma_z=0.002(1+z)$, \oiii{}-H$\beta$ interlopers (with a fraction of $5\%$), and the combination of the two, shown as the dashed, dot-dashed, and dotted lines, respectively. For comparison, the default case with RSD only is plotted as the black solid lines. The left and right panels show the power spectrum monopoles and quadrupoles, respectively. The upper (lower) panels show the results before (after) BAO reconstruction.}\label{fig:Pwnw_z0.6}
\end{figure*}

\subsection{BAO fitting model}
Given a pair of simulations, we calculate the wiggled and no-wiggle power spectra $P_{\text{wig}}(k,\mu)$ and $P_{\text{now}}(k,\mu)$, which contain and do not contain the BAO signal, respectively. The BAO power spectrum is extracted from the difference between $P_{\text{wig}}$ and $P_{\text{now}}$, i.e. $\delta P(k,\mu)=\Pwig - \Pnow$. We fit the BAO power spectrum with the widely used model \citep[e.g.][]{Seo2016,Beutler2017}, i.e.
\begin{align}
    \delta P_{\text{model}}(k, \mu) = b^2 (1+\beta\mu^2)^2 G_{\text{model}}^2(k, \mu)D^2_{\text{FoG}}(k, \mu)\delta P_{\text{lin}}(k), \label{eq:bao_fit_model}
\end{align}
where $\delta P_{\text{lin}}$ is the linear BAO power spectrum calculated from the difference between the input linear power spectra $P_{\text{wig}}$ and $P_{\text{now}}$. $G_{\text{model}}$ is given by equation (\ref{eq:model_prop}). $D_{\text{FoG}}$ denotes the Fingers of God (FoG) damping term, which is usually modelled as a Lorentzian function \citep[e.g.][]{Park1994,Ballinger1996}, i.e.
\begin{align}
    D_{\text{FoG}} = \frac{1}{1+k^2\mu^2 \Sigma^2_{\text{FoG}}/2}.
\end{align}
The FoG effect originates from the virial motions of galaxies in dark matter haloes \citep{Jackson1972}. Since we use halo catalogues instead of galaxy catalogues in this work, the FoG effect should be negligible. In addition, \cite{Chen_Howlett2024} claim that $D_{\text{FoG}}$ can be dropped without inducing noticeable systematics in BAO fitting even for DESI, which has strong statistical constraining power. In this study, we include this term to account for the additional damping effect not fully modelled by $\Sigmaperp$ and $\Sigmapara$, which are fixed in our fitting process. 

For the case of small-displacement interlopers, same as equation (\ref{eq:Pobs_int}), the modified BAO fitting model is
\begin{align}
    \delta P_\text{model}^{\text{int}} = \left[1 - 2I_\text{frac}(1- I_\text{frac})\big(1 - \text{cos}(kd^\text{int}\mu)\big)\right] \delta P_\text{model}.\label{eq:Pbao_int_model}
\end{align}
In Fig. \ref{fig:Pwnw_int_model_z0.6}, we show that the modified model matches well to the measurement.
The true (model) $(k',\mu')$ coordinates and the measured $(k, \mu)$ coordinates are related by $k'_{\perp}=k_{\perp}/\alpha_\perp$ and $k'_{\parallel}=k_{\parallel}/\alpha_{\parallel}$, i.e.
\begin{align}
    k'  &= \frac{k}{\alperp} \big[1+\mu^2(\alperp^2/\alpara^2 - 1)\big]^{1/2},\nonumber \\
    \mu'&=\mu \frac{\alperp}{\alpara}\big[1+\mu^2(\alperp^2/\alpara^2 - 1)\big]^{-1/2}, \label{eq:kmu_obs2true}
\end{align}
where $\alpara$ and $\alperp$ are the BAO dilation parameters parallel and perpendicular to the LoS, respectively. They are related to the cosmological distances and the intrinsic BAO scale by
\begin{align}
    \alpha_\perp(z) = \frac{D_\text{M}(z) r_\text{d}^{\text{fid}}}{D_\text{M}^{\text{\text{fid}}}(z)r_\text{d}},\;\;\;\;\;\;
    \alpha_\parallel(z) = \frac{D_\text{H}(z) r_\text{d}^{\text{fid}}}{D_\text{H}^{\text{\text{fid}}}(z)r_\text{d}},
\end{align}
where $D_\text{M}(z)$ is the comoving angular diameter distance, $D_\text{H}\equiv c/H(z)$ is the Hubble distance, and $r_\text{d}$ is the sound horizon at the baryon drag epoch. Since $\alperp$ and $\alpara$ are anti-correlated at $\sim 40$ per cent level, the less correlated reparameterizations $\alpiso$ and $\alpap$ are also often used, i.e.
\begin{align}
     \alpha_\text{iso} = (\alpha_\perp^2\alpha_\parallel)^{1/3},\;\;\;\;\; \alpha_{\text{AP}}=\alpha_\parallel/\alpha_\perp, \label{eq:alpiso_ap}
\end{align}
which characterize the isotropic shifting and anisotropic wraping of the BAO scale \citep{Padmanabhan2008}.

The modelled BAO power spectrum multipoles are given by
\begin{align}
    \delta P_{\ell}(k) = \frac{2\ell + 1}{2}\int_{-1}^{1}\delta P_\text{model}[k'(k,\mu), \mu'(\mu)]L_{\ell}(\mu)d\mu.
\end{align}
We fit the BAO power spectrum monopole $\delta P_0$ and quadrupole $\delta P_2$ over $k \in(0.02, 0.3)\hMpc$ with 56 uniform $k$ bins of width $0.005\hMpc$. We set $100$ uniform $\mu$ bins over $[0, 1.0]$ when calculating the anisotropic power spectrum.
Since we adopt the true cosmology for both measuring and modelling the BAO power spectra from the simulations, any departure of the fitted BAO scale parameters from unity indicates the presence of some systematics. 

\begin{figure*}
	\includegraphics[width=1.95\columnwidth]{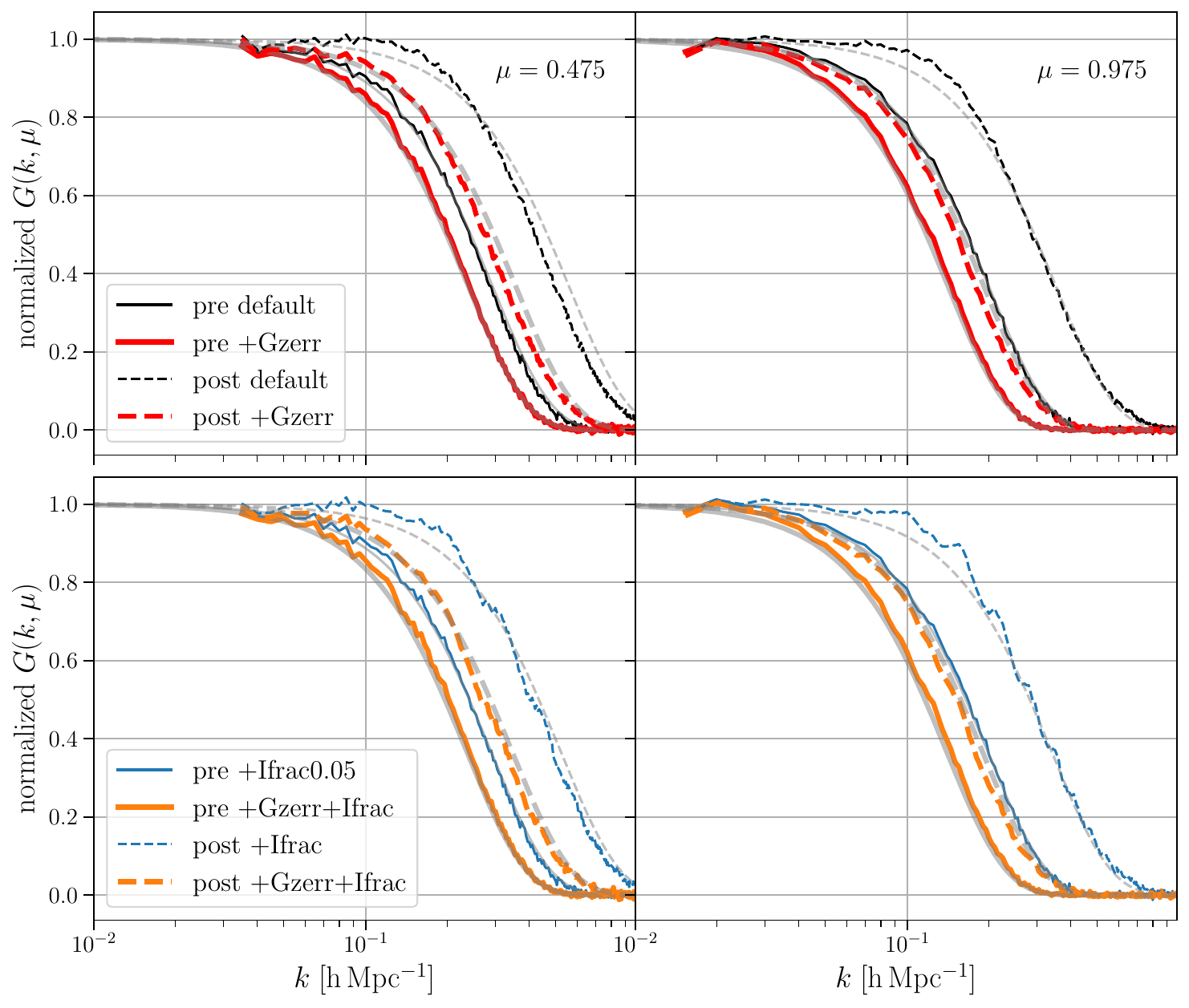}
    \caption{Effect of slitless spec-z uncertainty and interlopers on the propagators. \textit{Upper panels}: the propagators from the case with RSD only and the addition of spec-z uncertainty, shown as the thin black and thick red lines, respectively. The left and right panels show the propagators from the modes with $\mu=0.475$ and $\mu=0.975$, respectively. The BAO reconstruction can boost the propagators, shown as the dashed lines. The gray lines depict the theoretical prediction from equation (\ref{eq:model_prop}) given the estimated $\Sigma_{\perp}$ and $\Sigma_{\parallel}$. \textit{Lower panels}: same as the upper panels but for the case with the addition of small-displacement interlopers.}
    \label{fig:prop_z0.6}
\end{figure*}

\section{Results}\label{sec:results}
By varying the slitless spec-z errors in the mock catalogues, we investigate their effects on the BAO power spectrum, the propagator, and the BAO scale parameters. The notation used for each case is summarized in Table \ref{tab:notation} and adopted throughout the paper.
\begin{table*}
	\centering
	\caption{Notation for cases with different slitless spec-z  errors.}\label{tab:notation}
	\begin{tabular}{|c|c|}
		\hline
		notation & description  \\
		\hline
        default & RSD is the only systematic effect in the observed redshift\\
        \hline
        +Gzerr  & Default case plus Gaussian redshift uncertainty with $\sigma_z=0.002(1+z)$ \\
        \hline
        +Ifrac* & Default case plus small-displacement interlopers with fraction *\\
        \hline
        +Gzerr+Ifrac* & Default case plus both Gaussian redshift uncertainty and small-displacement interlopers with fraction *\\
		\hline
	\end{tabular}
\end{table*}

\subsection{BAO power spectrum}
Fig. \ref{fig:Pwnw_z0.6} compares the BAO power spectrum monopoles ($\ell=0$) and quadrupoles ($\ell=2$) with and without the slitless spec-z errors at $z=0.6$. We use \texttt{pypower}\footnote{\url{https://github.com/cosmodesi/pypower}} to calculate the power spectrum for each simulation. We obtain the BAO power spectrum by taking the difference between the power spectra of paired simulations. We compute the mean BAO power spectra over 90 realizations before and after BAO reconstruction, shown in the upper and lower panels, respectively. The left and right columns show the monopoles and quadrupoles, respectively. The black solid lines represent the default case. In addition to the default, the remaining lines correspond to cases with additional spec-z errors, i.e. the Gaussian redshift uncertainty with $\sigma_0=0.002$ (blue dashed lines); the small-displacement interlopers with a $5\%$ fraction (orange dot-dashed lines), and the combination of both (green dotted lines). Both the spec-z uncertainty and interlopers damp the BAO signal. The redshift uncertainty shifts the phase of BAO signal in the quadrupole at $k>0.1\hMpc$. BAO reconstruction significantly enhances the BAO signal, even in the presence of spec-z errors.

From the BAO power spectrum, neither the Gaussian spec-z uncertainty nor the small-displacement interlopers appear to shift the BAO peak position significantly, though the broadband shape can be substantially affected, as shown in Fig. \ref{fig:pkmu_now_z0.6}. The Gaussian redshift uncertainty significantly damps the broadband shape. The small-displacement interlopers not only damp the power spectrum amplitude, but also induce significant oscillations in the power spectrum shape, as also shown in \citep{Cagliari2025}.  

\subsection{Propagator and nonlinear BAO damping parameters}\label{sec:propagator_result}
\begin{table*}
	\centering
	\caption{Nonlinear BAO damping parameters $\Sigmaperp$ and $\Sigmapara$ estimated from the propagators. The default case includes only RSD. +Gzerr and +Ifrac denote the addition of Gaussian redshift uncertainty and small-displacement interlopers, respectively. The spec-z uncertainty is set to $\sigma_z=0.002(1+z)$. The number following +Ifrac denotes the interloper fraction. We compare the cases before and after BAO reconstruction, and show the post-reconstruction results in the parentheses.}
	\label{tab:Sigma_nl}
	\begin{tabular}{|l|c|c|c|c|c|c|c|c|} 
		\hline
        redshift & default & +Gzerr & +Ifrac0.01 & +Ifrac0.05 & +Ifrac0.1 & +Gzerr+Ifrac0.01 & +Gzerr+Ifrac0.05 & +Gzerr+Ifrac0.1 \\  
        \hline
        & \multicolumn{8}{c}{$\Sigmaperp\;[\Mpch]$}\\ \hline
$0.0$ & $7.3\, (3.6)$ & $7.3\, (3.7)$ & $7.3\, (3.6)$ & $7.4\, (3.9)$ & $7.4\, (4.1)$ & $7.3\, (3.8)$ & $7.4\, (4.0)$ & $7.4\, (4.2)$ \\$0.6$ & $5.5\, (2.6)$ & $5.5\, (2.7)$ & $5.5\, (2.7)$ & $5.6\, (2.9)$ & $5.6\, (3.2)$ & $5.5\, (2.8)$ & $5.6\, (3.0)$ & $5.6\, (3.3)$ \\$1.0$ & $4.4\, (1.9)$ & $4.4\, (2.1)$ & $4.4\, (2.0)$ & $4.5\, (2.3)$ & $4.6\, (2.6)$ & $4.4\, (2.1)$ & $4.5\, (2.4)$ & $4.6\, (2.7)$ \\\hline 
& \multicolumn{8}{c}{$\Sigmapara\;[\Mpch]$}\\ \hline 
$0.0$ & $11.6\, (6.3)$ & $14.1\, (10.2)$ & $11.6\, (6.3)$ & $11.6\, (6.4)$ & $11.5\, (6.5)$ & $14.0\, (10.2)$ & $13.9\, (10.2)$ & $13.8\, (10.3)$ \\$0.6$ & $10.2\, (5.8)$ & $14.5\, (10.7)$ & $10.2\, (5.8)$ & $10.3\, (6.0)$ & $10.4\, (6.3)$ & $14.6\, (10.7)$ & $14.6\, (10.8)$ & $14.6\, (11.0)$ \\$1.0$ & $8.9\, (5.2)$ & $13.2\, (10.8)$ & $8.9\, (5.2)$ & $9.0\, (5.2)$ & $9.1\, (5.4)$ & $13.1\, (10.8)$ & $12.9\, (10.8)$ & $12.8\, (10.9)$ \\
		\hline
	\end{tabular}
\end{table*}

We compute the propagators using equation (\ref{eq:propagator}) with $\mu$ bins of width $0.05$. Fig. \ref{fig:prop_z0.6} shows the measured propagators $G(k)$ at $z=0.6$ with and without spec-z errors. We show the mean propagators averaged over 10 realizations to reduce statistical noise. The left and right panels show the results from the $\mu$ bins centered at $0.475$ and $0.975$, respectively. 
In the upper panels, the thin black lines depict the default case, compared to the case with additional Gaussian spec-z uncertainty, shown as the thick red lines. The redshift uncertainty clearly damps $G(k)$. Solid and dashed lines correspond to the results before and after BAO reconstruction, labeled `pre' and `post', respectively. After reconstruction, $G(k)$ is enhanced and approaches unity, even for the case with Gaussian spec-z uncertainty. This demonstrates that BAO reconstruction remains effective for CSST slitless spec-z samples with a redshift uncertainty of $\sim 0.002(1+z)$.  

We estimate the nonlinear BAO damping parameters $\Sigmaperp$ and $\Sigmapara$ from the measured $G(k)$. At $k_p=0.3\hMpc$, we match the model (equation \ref{eq:model_prop}) to the measured propagators at $\mu=0$ and $1.0$, and obtain $\Sigmaperp$ and $\Sigmapara$, respectively. Given $\Sigmaperp$ and $\Sigmapara$, we compute the modelled $G(k)$ and plot them as the gray lines. The line styles and thickness are matched to those of the corresponding observed lines. Overall, the model predictions agree well with the measurements. 

Similarly, in the lower panels, we show the results from the cases with the addition of small-displacement interlopers (with a $5\%$ interloper fraction). We also consider the cases with and without spec-z uncertainty. As with the power spectrum, small-displacement interlopers both damp the amplitude and induce oscillations in $G(k)$. We correct this effect by accounting for the factor $1-I_{\text{frac}}\big(1-\cos(kd^\text{int}\mu)\big)$. Without correction, it can bias the estimates of $\Sigmaperp$ and $\Sigmapara$. For the pre-reconstruction case, our model (equation \ref{eq:model_prop_int}) yields $\Sigmaperp$ and $\Sigmapara$ consistent with those of the default case. For the post-reconstruction case, there are some residual oscillations not fully corrected by the model.

Beyond $z=0.6$, we estimate $\Sigmaperp$ and $\Sigmapara$ from the propagators at $z=0$ and $1.0$ as well, and show the results in Table \ref{tab:Sigma_nl}. Compared to the default case, the Gaussian redshift uncertainty increases $\Sigmapara$ by $3\sim4\Mpch$ and $4\sim 6\Mpch$, depending on redshift, for the pre- and post-reconstruction cases, respectively. The redshift uncertainty can mimic the effect of nonlinear structure evolution and RSD, smearing the BAO signal along the LoS.

For small-displacement interlopers, we also study cases with lower and higher interloper fractions, i.e. $1\%$ and $10\%$. After accounting for the interloper effect in the model, we get unbiased $\Sigmaperp$ and $\Sigmapara$ for the pre-reconstruction case. For the post-reconstruction, $\Sigmaperp$ and $\Sigmapara$ are slightly overestimated, e.g. by $\sim 0.5\Mpch$ for the 10\% interloper fraction case, with a smaller bias for lower fractions.

\subsection{BAO scale parameters}
\begin{table*}
	\centering
	\setlength{\tabcolsep}{3pt} 
	\caption{The mean BAO scale parameters $\alperp$ and $\alpara$ fitted from 90 realizations, with the standard deviations ($68\%$ confidence level) of the means. We show the deviation of $\alpha$ from unity as a percentage. We compare results for cases with different spec-z errors, using the same notation as in Table \ref{tab:Sigma_nl}. The upper and lower parts show the results before and after BAO reconstruction, labeled `pre-recon' and `post-recon', respectively.}\label{tab:bao_params_sigma0.002}
	\begin{tabular}{|l|c|c|c|c|c|c|c|c|} 
		\hline
        redshift & default & +Gzerr & +Ifrac0.01 & +Ifrac0.05 & +Ifrac0.1 & +Gzerr+Ifrac0.01 & +Gzerr+Ifrac0.05 & +Gzerr+Ifrac0.1 \\ \hline 
        & \multicolumn{8}{c}{$\alperp-1$ [\%] (pre-recon)} \\ \hline
0.0 & $0.108 \pm 0.019 $ & $0.095 \pm 0.026 $ & $0.127 \pm 0.023 $ & $0.135 \pm 0.032 $ & $0.136 \pm 0.040 $ & $0.105 \pm 0.028 $ & $0.107 \pm 0.037 $ & $0.103 \pm 0.044 $ \\ \hline
0.6 & $0.128 \pm 0.018 $ & $0.124 \pm 0.029 $ & $0.120 \pm 0.021 $ & $0.150 \pm 0.030 $ & $0.187 \pm 0.037 $ & $0.116 \pm 0.032 $ & $0.143 \pm 0.039 $ & $0.195 \pm 0.046 $ \\ \hline
1.0 & $0.209 \pm 0.023 $ & $0.226 \pm 0.043 $ & $0.204 \pm 0.031 $ & $0.195 \pm 0.042 $ & $0.126 \pm 0.061 $ & $0.224 \pm 0.047 $ & $0.213 \pm 0.060 $ & $0.140 \pm 0.068 $ \\ \hline
& \multicolumn{8}{c}{$\alpara-1$ [\%] (pre-recon)} \\ \hline
0.0 & $0.495 \pm 0.032 $& $0.550 \pm 0.074 $& $0.491 \pm 0.040 $& $0.422 \pm 0.049 $& $0.439 \pm 0.059 $& $0.573 \pm 0.073 $& $0.556 \pm 0.081 $& $0.641 \pm 0.093 $\\ \hline
0.6 & $0.528 \pm 0.035 $& $0.769 \pm 0.097 $& $0.480 \pm 0.037 $& $0.452 \pm 0.051 $& $0.377 \pm 0.067 $& $0.726 \pm 0.097 $& $0.747 \pm 0.110 $& $0.650 \pm 0.133 $\\ \hline
1.0 & $0.605 \pm 0.041 $& $0.437 \pm 0.155 $& $0.571 \pm 0.059 $& $0.592 \pm 0.083 $& $0.658 \pm 0.114 $& $0.446 \pm 0.166 $& $0.605 \pm 0.191 $& $0.697 \pm 0.221 $\\ \hline
& \multicolumn{8}{c}{$\alperp-1$ [\%] (post-recon)} \\ \hline
0.0 & $0.057 \pm 0.011 $ & $0.073 \pm 0.017 $ & $0.054 \pm 0.013 $ & $0.040 \pm 0.019 $ & $0.028 \pm 0.023 $ & $0.070 \pm 0.019 $ & $0.055 \pm 0.026 $ & $0.025 \pm 0.028 $ \\ \hline
0.6 & $0.026 \pm 0.013 $ & $0.038 \pm 0.022 $ & $0.024 \pm 0.015 $ & $0.056 \pm 0.022 $ & $0.044 \pm 0.031 $ & $0.025 \pm 0.025 $ & $0.044 \pm 0.034 $ & $0.051 \pm 0.040 $ \\ \hline
1.0 & $0.001 \pm 0.017 $ & $0.021 \pm 0.038 $ & $-0.003 \pm 0.023 $ & $0.006 \pm 0.039 $ & $-0.023 \pm 0.052 $ & $0.007 \pm 0.042 $ & $0.015 \pm 0.052 $ & $-0.003 \pm 0.061 $ \\ \hline
& \multicolumn{8}{c}{$\alpara-1$ [\%] (post-recon)} \\ \hline
0.0 & $0.114 \pm 0.018 $& $0.207 \pm 0.046 $& $0.106 \pm 0.022 $& $0.098 \pm 0.032 $& $0.069 \pm 0.041 $& $0.203 \pm 0.048 $& $0.214 \pm 0.061 $& $0.216 \pm 0.071 $\\ \hline
0.6 & $0.067 \pm 0.020 $& $0.159 \pm 0.074 $& $0.048 \pm 0.024 $& $0.010 \pm 0.034 $& $-0.004 \pm 0.047 $& $0.125 \pm 0.076 $& $0.134 \pm 0.087 $& $0.089 \pm 0.112 $\\ \hline
1.0 & $0.059 \pm 0.026 $& $-0.020 \pm 0.118 $& $0.049 \pm 0.038 $& $0.049 \pm 0.065 $& $0.066 \pm 0.089 $& $-0.015 \pm 0.126 $& $0.036 \pm 0.142 $& $0.146 \pm 0.170 $\\ \hline
		\hline
	\end{tabular}
\end{table*}

In the BAO fitting model (equation \ref{eq:bao_fit_model}), we treat $\alperp$, $\alpara$, $b$, $f$, $\Sigma_{\text{FoG}}$ as free parameters. We set uniform priors on $\alperp\in(0.8, 1.2)$, $\alpara\in(0.8, 1.2)$, $b\in(0.2, 3.0)$, $f\in(0.2, 1.5)$ and $\Sigma_{\text{FoG}}\in(0, 15)\Mpch$. We fix $\Sigmaperp$ and $\Sigmapara$ to the values estimated from the measured propagators, as shown in Table \ref{tab:Sigma_nl}\footnote{We have checked that leaving $\Sigmaperp$ and $\Sigmapara$ free rather than fixing them has little effect on the best-fit $\alperp$ and $\alpara$.}. For interloper cases, we use the modified fitting model from equation~(\ref{eq:Pbao_int_model}). Since $\Sigmaperp$ and $\Sigmapara$ do not change much, we set them to be the same values as in the default or +Gzerr cases, depending on the addition of Gaussian spec-z uncertainty or not. In addition, we fix the interloper fraction and displacement to their true values. 

\begin{figure*}
    \centering
    \includegraphics[width=0.495\linewidth]{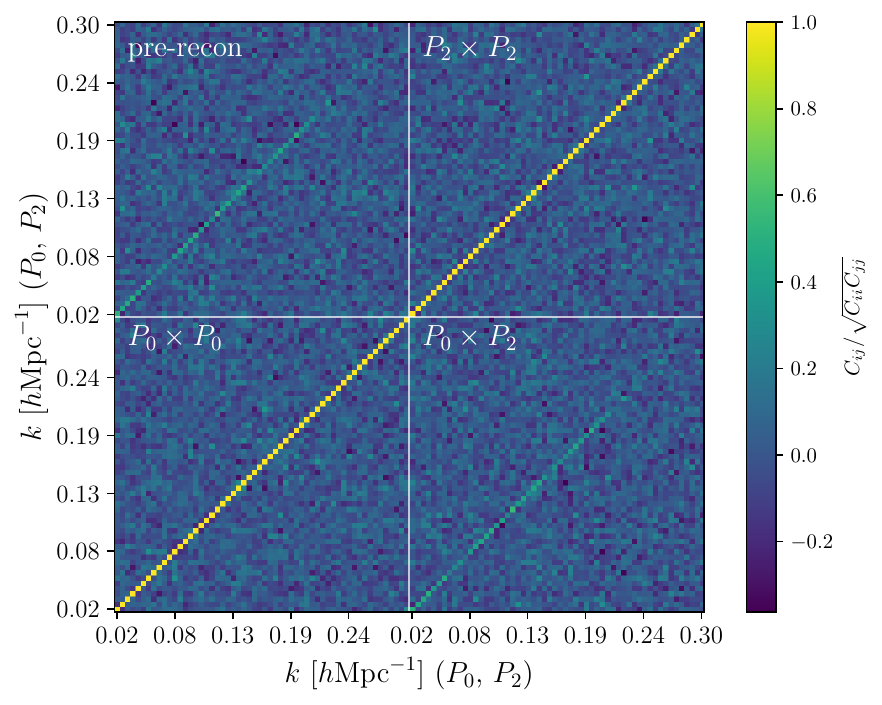}
    \includegraphics[width=0.495\linewidth]{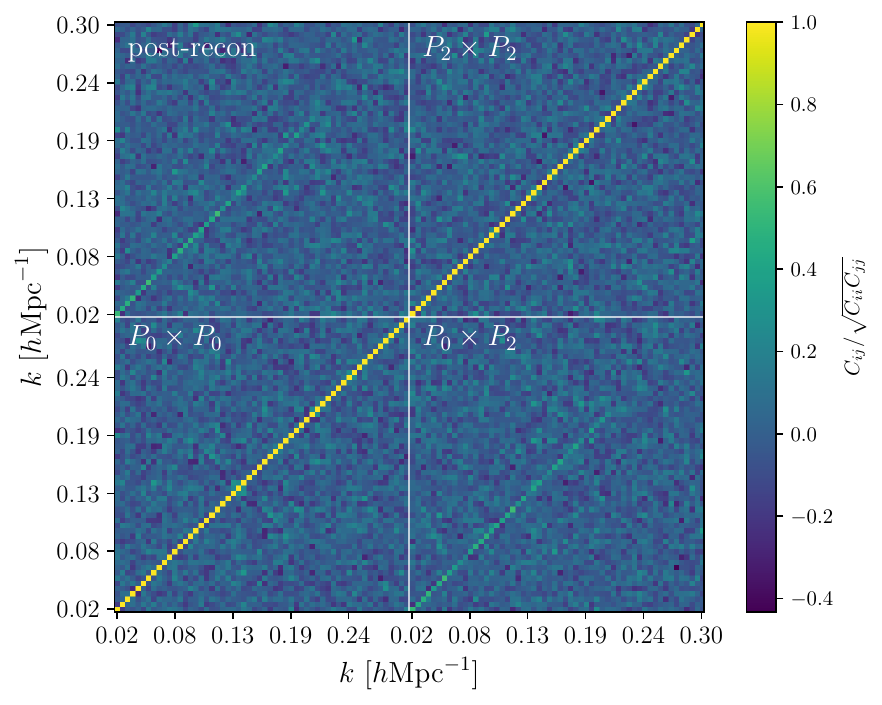}
    \caption{Correlation coefficients of the BAO power spectrum covariance matrix for the +Gzerr+Ifrac0.05 case at $z=0.6$. The left and right panels show the results before and after the BAO reconstruction, respectively. Each matrix contains the monopole (bottom-left corner) and quadrupole (top-right corner), as well as their cross-correlation (top-left and bottom-right). Orange and blue denote positive and negative correlations, respectively.}
    \label{fig:cov}
\end{figure*}

We compute the covariance matrix of the BAO power spectrum monopole and quadrupole from $90$ paired simulations, i.e.
\begin{align}
    \text{Cov}(\delta P_{\ell}(k_i), \delta P_{\ell'}(k_j)) =\frac{1}{N-1} \sum_{n=1}^{N} (\delta P_{\ell i,n} - \overline{\delta P_{\ell i}})(\delta P_{\ell' j,n} - \overline{\delta P_{\ell' j}}), \label{eq:cov_bao_pk}
\end{align}
where $\delta P_{\ell i,n}$ denotes the BAO power spectrum multipole at $k_i$ from the $n$th pair of simulations, and the overline denotes the mean over all realizations. The correlation coefficients of the covariance matrix are defined as 
\begin{align}
r_{ij} =  \frac{C_{ij}}{\sqrt{C_{ii} C_{jj}}},  
\end{align}
which characterize the cross-correlation between different bins.
In Fig. \ref{fig:cov}, we plot the correlation coefficients for the +Gzerr+Ifrac0.05 case at $z=0.6$ before and after reconstruction. As shown, the diagonal terms dominate with some cross-correlation between the monopole and quadrupole.

Since the number of simulations is smaller than the number of $k$ bins, i.e. 90 versus 112, we cannot accurately invert the covariance matrix \citep{Hartlap2007, Percival2014}. Instead, assuming a Gaussian covariance, we consider only the diagonal terms of Cov$(\delta P_0,\, \delta P_0)$, Cov$(\delta P_0,\, \delta P_2)$, and Cov$(\delta P_2,\, \delta P_2)$, and obtain the inverse covariance for BAO fitting. 
To mitigate the effect of the Gaussian covariance assumption, we derive the errors on the BAO parameters from the dispersion across the 90 individual best-fits, rather than from fitting the mean power spectrum (see \citealt{Schmittfull2015, Schmittfull2017} for the same approach).
In addition, since we are primarily interested in systematic bias on the BAO parameters from the spec-z errors, we focus on the relative difference with respect to the default case, which should be relatively insensitive to the adopted covariance matrix. 

We perform the BAO fitting by maximizing the Gaussian likelihood, i.e. minimizing the $\chi^2$:
\begin{align}
    \chi^2 = (\delta P_{\text{obs}} -\delta P_{\text{model}})^T \operatorname{Cov}^{-1}(\delta P_{\text{obs}} -\delta P_{\text{model}}). 
\end{align}
We fit parameters using the affine-invariant Markov chain Monte Carlo sampler \citep{Goodman2010}, implemented in \texttt{emcee}\footnote{\url{https://github.com/dfm/emcee}}\citep{2013PASP..125..306F}. We use 64 walkers with a maximum of 20,000 steps for the chain to ensure fitting convergence. To monitor convergence, we compute the integrated autocorrelation time of the chain every 100 sampling steps. If the total number of sampling steps exceeds 100 times the estimated autocorrelation time, and the relative variation of the autocorrelation time between consecutive evaluations is less than 1\%, we deem the fit converged\footnote{An example implementation using the autocorrelation time as a convergence criterion is available at \url{https://emcee.readthedocs.io/en/v3.0.0/tutorials/monitor/.}}.
For each simulation pair, we fit the BAO power spectrum monopole and quadrupole to obtain the marginalized mean of $\alperp$ and $\alpara$. Fig. \ref{fig:alperp_para_z1.0} shows an example of the $\alperp$ and $\alpara$ best-fit values from 40 (out of 90) realizations at $z=1.0$.

Table \ref{tab:bao_params_sigma0.002} shows the mean of $\alperp$ and $\alpara$ best-fit values over 90 realizations for different cases. The results before and after BAO reconstruction are shown in the upper and lower halves of the table, respectively. 

\begin{figure*}
	\includegraphics[width=1.95\columnwidth]{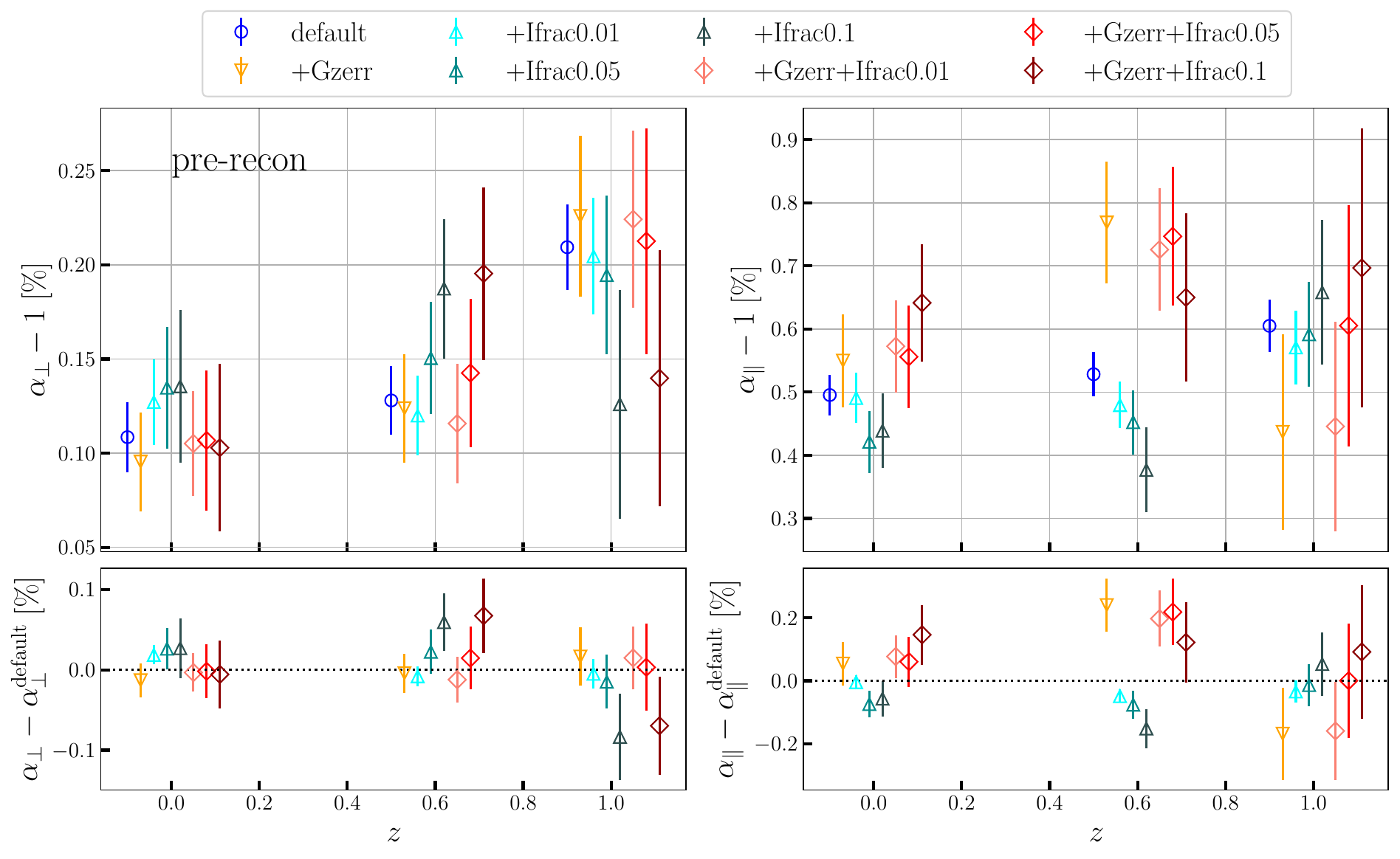}
    \caption{\textit{Upper panels}: the mean of the $\alperp$ (left) and $\alpara$ (right) best-fit values from 90 realizations. Circles denote the default case. Downward-pointing triangles denote the case with additional spec-z uncertainty $\sigma_z=0.002(1+z)$, and upward-pointing triangles denote the case with interlopers. Diamonds denote the case with both spec-z uncertainty and interlopers. Within each marker type, darker shading indicates a higher interloper fraction. \textit{Lower panels:} the mean difference in $\alperp$ and $\alpara$ between the spec-z error cases and the default case, averaged over all realizations. Most points lie within $\sim 2\sigma$ of zero, indicating no significant bias from the spec-z errors.}
    \label{fig:alphas_mean_pre}
\end{figure*}

\begin{figure*}
	\includegraphics[width=1.95\columnwidth]{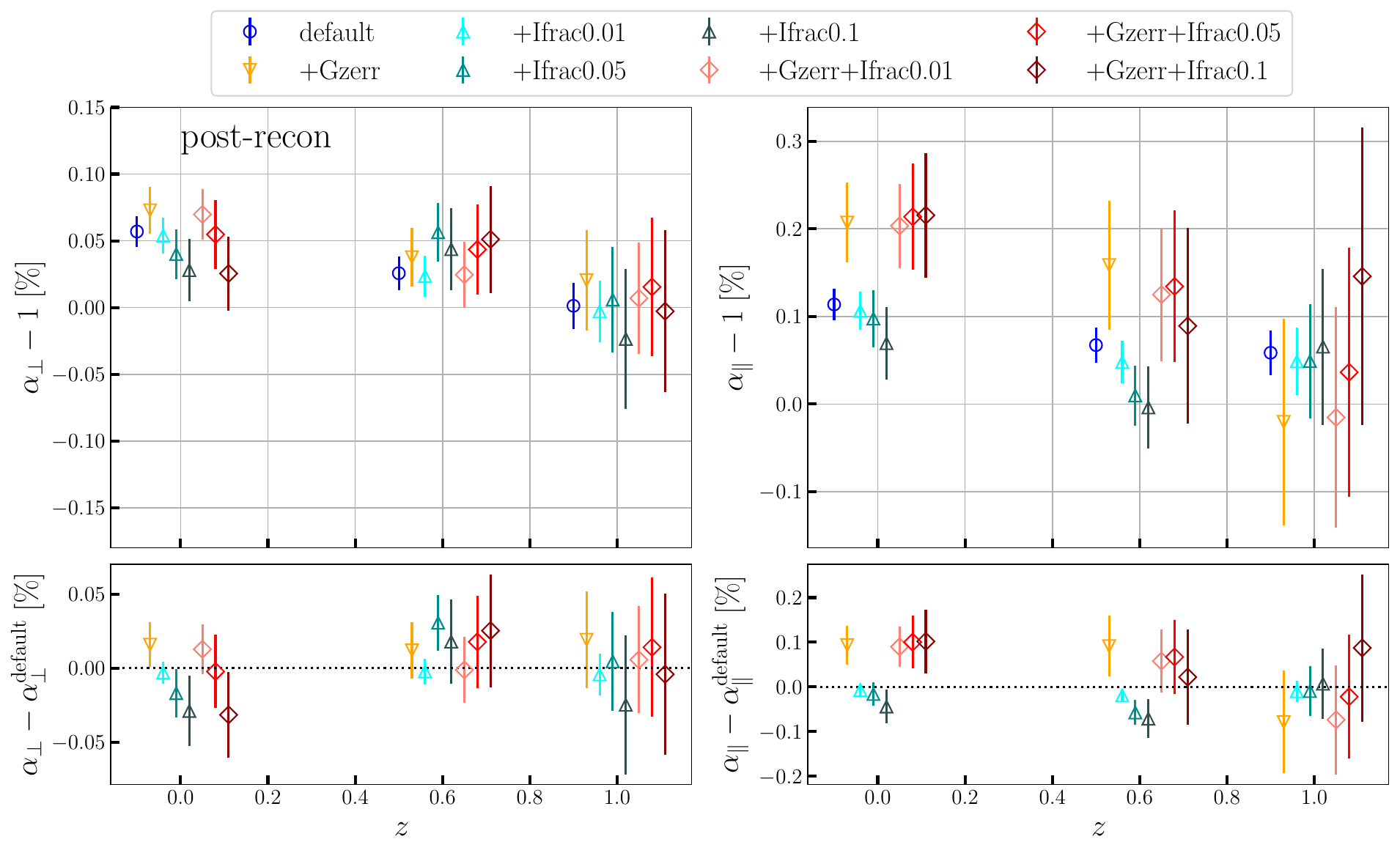}
    \caption{Same as Fig. \ref{fig:alphas_mean_pre} but for the post-reconstruction results.}
    \label{fig:alphas_mean_post}
\end{figure*}
Correspondingly, the means and standard deviations of the best-fit $\alpha$ parameters over all realizations are shown in Figs. \ref{fig:alphas_mean_pre} and \ref{fig:alphas_mean_post} for the pre- and post-reconstruction cases, respectively. In each figure, the upper panels show the mean best-fit values for different cases. The left and right panels show $\alperp$ and $\alpara$, respectively. Circles denote the default case. Downward-pointing triangles denote the case with additional Gaussian spec-z uncertainty ($\sigma_0=0.002$). 
Upward-pointing triangles denote the case with interlopers, with darker shading indicating higher interloper fractions. Diamonds denote the case with both Gaussian spec-z uncertainty and interlopers. 

For each case with slitless spec-z errors, we compute the relative difference in $\alpha$ best-fit values from the default case for each realization. We obtain the mean difference and its standard deviation over all realizations. The results are shown in the lower panels of Figs. \ref{fig:alphas_mean_pre} and \ref{fig:alphas_mean_post}. At a given redshift, most data points cluster within $\sim 2 \sigma$ of zero, indicating no significant systematic bias in the BAO signal from slitless spec-z errors after correct modelling. Given the statistical constraints, we conclude that the systematic biases of $\alperp$ and $\alpara$ are controlled to below $0.1$ per cent and $0.2$ per cent, respectively, for the spec-z error cases studied here. This conclusion holds for both pre- and post-reconstruction.

Reconstruction helps reduce the BAO systematics induced by nonlinear structure growth and RSD. For the default case, reconstruction reduces the bias in $\alperp$ from above $0.1$ per cent to below $0.1$ per cent, and the bias in $\alpara$ from above $0.5$ per cent to below $0.2$ per cent. It also significantly reduces the scatter in the $\alpha$ best-fit values. This behavior is similar for cases with slitless spec-z errors. Compared to the default case, the systematic bias from the spec-z errors is smaller after reconstruction, i.e. $\lesssim 0.05$ per cent and $\lesssim 0.1$ per cent for $\alperp$ and $\alpara$, respectively.

Compared to the impact of the interlopers, the Gaussian spec-z substantially increases the errors on $\alpha$, particularly at higher redshifts, indicating that it is a main factor limiting the precision of BAO scale measurement from the CSST slitless spec-z survey. We present a more detailed investigation of this effect in Sec. \ref{sec:diff_Gzerr}, where we consider larger Gaussian spec-z uncertainties.

For each realization, we take the marginalized means of $\alperp$ and $\alpara$, and derive the corresponding $\alpiso$ and $\alpap$ using equation (\ref{eq:alpiso_ap}). We then compute the mean and standard deviation across the 90 realizations. We show the results of $\alpiso$ and $\alpap$ in Figs.~\ref{fig:alpiso_ap_prerecon} and ~\ref{fig:alpiso_ap_postrecon} for the pre- and post-reconstruction cases, respectively. The systematic effect of spec-z errors on $\alpiso$ and $\alpap$ is similar to that of $\alperp$ and $\alpara$.

\subsection{Varying Gaussian redshift uncertainty}\label{sec:diff_Gzerr}
In addition to the default spec-z uncertainty $\sigma_0=0.002$, we consider more pessimistic cases with $\sigma_0=0.004$ and $0.006$. We adopt the same halo mass cut as in the default case; hence, the sample size remains unchanged as $\sigma_0$ increases. In a real survey, a larger spec-z uncertainty would be accompanied by a larger sample size owing to looser data selection criteria. Such an effect may affect our result at high redshift, where the sample density is relatively low. We leave it for future investigation. Fig.~\ref{fig:bao_z0.6_diffGzerr} shows the BAO signal for cases with different amounts of redshift uncertainty for both the pre- and post-reconstruction cases. We perform the same analyses as above and estimate $\Sigmaperp$ and $\Sigmapara$ from the propagators, as shown in Fig.~\ref{fig:Sigmanl_diffGzerr}. 
\begin{figure*}
  \includegraphics[width=1.95\columnwidth]{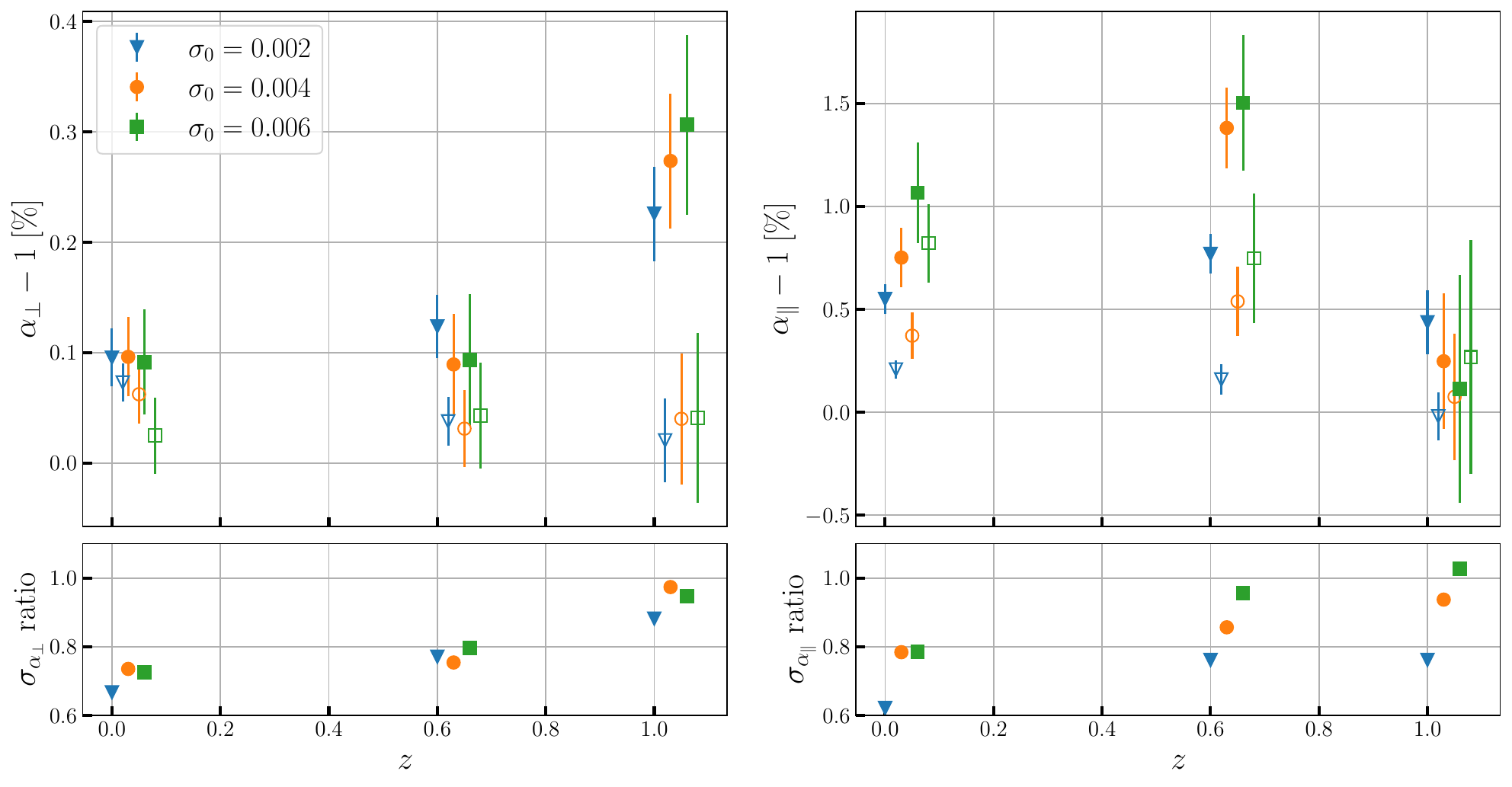}
    \caption{$\alperp$ (left) and $\alpara$ (right) obtained from fits to the cases with different spec-z uncertainties. \textit{Upper panels:} the mean best-fit values across the 90 realizations, with error bars representing the standard deviations of the means. Solid and empty points correspond to the results before and after BAO reconstruction, respectively. \textit{Lower panels:} the ratio of the post-reconstruction to pre-reconstruction standard deviations of $\alpha$. A larger ratio indicates a smaller gain from reconstruction.}
    \label{fig:alphas_diffGzerr}
\end{figure*}

Following the same procedure as above, we fit the BAO power spectrum multipoles and obtain the BAO scale parameters for cases with different spec-z uncertainties. For each realization, we obtain the marginalized means of $\alperp$ and $\alpara$ from the fit, then compute the mean and standard deviation across the 90 realizations. We show the results in the upper panels of Fig. \ref{fig:alphas_diffGzerr}. The left and right panels show $\alperp$ and $\alpara$, respectively. Different markers represent different spec-z uncertainties. We compare the results before and after reconstruction, shown as solid and empty points, respectively. We perform BAO reconstruction using the same smoothing scale of $15\Mpch$, which could be further optimized as the spec-z uncertainty and the tracer number density vary. At each redshift, data points are offset horizontally by a small amount to avoid overcrowding. Overall, reconstruction improves the BAO scale measurement, yielding less systematic bias. At $z=1.0$, for $\sigma_0=0.004$ and $\sigma_0=0.006$, although reconstruction yields little improvement on $\alpara$, it helps reduce the bias in $\alperp$.

In the lower panels, we show the ratio of the post-reconstruction to pre-reconstruction errors on $\alpha$. A larger value indicates a smaller gain from reconstruction. At $z=1.0$ for $\sigma_0=0.004$ and $\sigma_0=0.006$, reconstruction yields no improvement in the precision of $\alpara$. For $\alperp$, reconstruction reduces the errors by $\gtrsim 20$ per cent at $z=0$ and $0.6$, but the reduction is smaller at $z=1.0$. This difference is likely due to the tracer number density, since the uncertainties in the comoving distance associated with a given spec-z uncertainty are similar at the three redshifts. Recently, \cite{Chan2024} proposed a 2D BAO reconstruction scheme based on the projected density field from a photometric redshift survey, for which the redshift uncertainty is much larger than that of a slitless spec-z survey. Their method can enhance the BAO signal from the modes perpendicular to the LoS for dense tracer samples. Our results seem consistent with their findings.

\section{Conclusions and discussions}
\label{sec:conclusion}
We study the effect of slitless spectroscopic redshift uncertainty and interlopers on BAO measurements from CSST-like samples.
We construct CSST-like galaxy samples from \textsc{FastPM} dark matter haloes by selecting the most massive haloes, matching the galaxy number density and bias to values expected from the slitless spec-z survey. We apply the sample variance cancellation technique to the paired simulations. Even with only 90 realizations, the statistical constraining power on BAO is considerable.
We consider three redshift snapshots, $z=0.0$, $0.6$ and $1.0$, representative of the spec-z range (z<1) that CSST will cover.
Apart from the default case with RSD only, we consider redshift uncertainty and \oiii{}-$\Hbeta$ small-displacement interlopers as sources of slitless spec-z errors. 
In addition, we apply BAO reconstruction to the mocks and study the effect of spec-z errors on the reconstructed results. 

We model the spec-z uncertainty as Gaussian, with a default value of $\sigma_z=0.002(1+z)$. We find that it significantly damps the broadband shape of the power spectrum, as well as the BAO signal in the monopole and quadrupole.
We study its impact on the propagator and estimate the nonlinear BAO damping parameters $\Sigmaperp$ and $\Sigmapara$ from it. With the default spec-z uncertainty, $\Sigmapara$ increases by $3\sim 4\Mpch$ before reconstruction and by $4\sim 6\Mpch$ after reconstruction, with some dependence on redshift, while $\Sigmaperp$ changes little.  
For interloper case, we vary the interloper fraction $I_{\text{frac}}$, adopting values of $1\%$, $5\%$ and $10\%$. We consider the effect of interlopers on the observed overdensity contrast, and derive the corrected propagator model.
We obtain unbiased estimates of $\Sigmaperp$ and $\Sigmapara$ before BAO reconstruction, and slight overestimation (by $\sim 0.5\Mpch$) after reconstruction.

We fit the BAO power spectrum without modelling the broadband shape, and obtain the anisotropic BAO scale parameters $\alperp$ and $\alpara$. Both the redshift uncertainty and small-displacement interlopers can damp the BAO amplitude. We derive the BAO fitting model in the presence of small-displacement interlopers. From these fits, we compare the $\alpha$ values obtained with slitless spec-z errors to those of the default case, and confirm that there is no significant systematic bias. The relative differences in $\alperp$ and $\alpara$ from the default case are mostly within $0.1$ per cent
and $0.2$ per cent, respectively, for both pre- and post-reconstruction. The same holds for $\alpiso$ and $\alpap$. With the addition of spec-z errors, the statistical errors on $\alpha$ increase. The impact of redshift uncertainty is larger than that of the interlopers, especially at higher redshifts. In addition, we find that BAO reconstruction is necessary for CSST slitless spec-z samples with the default redshift uncertainty, since the systematic bias and statistical errors on $\alpha$s can be significantly reduced by reconstruction.

Apart from the default redshift uncertainty, we consider more pessimistic cases with $\sigma_0=0.004$ and $0.006$. We study the fitted BAO scale parameters before and after reconstruction. Overall, reconstruction can still decrease the bias and statistical errors on $\alpha$, e.g. reducing the bias in $\alperp$ at $z=1.0$, as well as the errors of $\alperp$ by $\gtrsim 20$ per cent at $z=0.0$ and $0.6$. Owing to the relatively low number density, reconstruction improves little in the precision of the BAO measurement at $z=1.0$.
Overall, our work provides a reference for CSST BAO measurements using real data, as well as for other slitless spectroscopic redshift surveys.

\section*{Acknowledgements}
We thank Hui Peng, Run Wen, Yu Liu, and Zhongxu Zhai for helpful discussions. We acknowledge the anonymous referee for constructive suggestions to improve this manuscript. We used Doubao, a large language model, for linguistic polishing. This work is supported by the National Key R\&D Program of China (Grant No. 2023YFA1607800, No. 2023YFA1607801, No. 2023YFA1607802, No. 2025YFA1614103), the National Natural Science Foundation of China (Grant No. 12595311, No. 12273020, No. 12503112), the China Manned Space Project with No. CMS-CSST-2025-A04.
Z. D. and Y.Y. acknowledge the sponsorship from Yangyang Development Fund. This work makes use of the Gravity Supercomputer at the Department of Astronomy, Shanghai Jiao Tong University.

\section*{Data Availability}
Data and Python scripts producing the figures in this manuscript can be found at \url{https://doi.org/10.5281/zenodo.20619777}.

\bibliographystyle{mnras}
\bibliography{references}

\appendix

\section{Supplementary figures}
In this section, we present the supplementary figures.

\begin{figure*}
	\includegraphics[width=1.95\columnwidth]{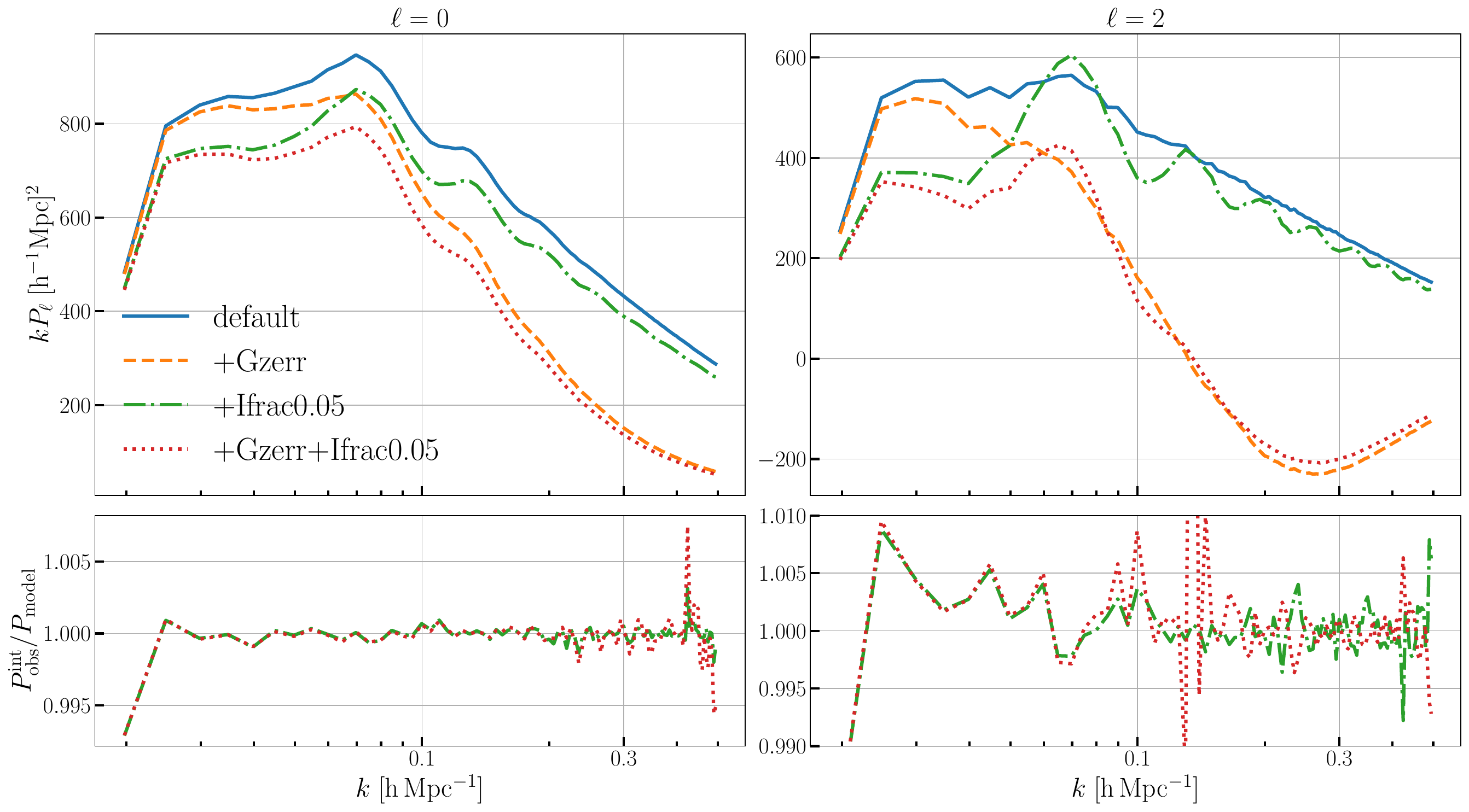}
    \caption{\textit{Upper panels:} effect of slitless spec-z errors on the broadband shape of the power spectrum monopole (left) and quadrupole (right) at $z=0.6$. We adopt a Gaussian spec-z uncertainty $\sigma_0=0.002$, and an interloper fraction of $5\%$. We show the mean power spectra (including the BAO signal) averaged over 90 realizations. \textit{Lower panels:} ratio of the observed power spectra to the model prediction for the interloper cases. The dashed and dotted lines correspond to the +Ifrac$0.05$ and +Gzerr+Ifrac$0.05$ cases, respectively.}
    \label{fig:pkmu_now_z0.6}
\end{figure*}
In the upper panels of Fig. (\ref{fig:pkmu_now_z0.6}), we show the effects of different slitless spec-z errors on the broadband shape of the power spectrum monopole (left) and quadrupole (right). We plot the mean multipoles averaged over 90 realizations at $z=0.6$. Different line types represent the cases with different spec-z errors. The blue solid lines correspond to the default case. The addition of the Gaussian spec-z uncertainty ($\sigma_0=0.002$) significantly damps the broadband shape, as shown by the orange dashed lines. The addition of small-displacement interlopers not only damps the amplitude of the power spectrum but also induces oscillations, especially in the quadrupole, shown as the green dot-dashed lines. 

In the lower panels of Fig. (\ref{fig:pkmu_now_z0.6}), we show the ratio of the mean observed power spectra to the model prediction in the presence of small-displacement interlopers. Based on equation (\ref{eq:Pobs_int}), we model the +Ifrac$0.05$ power spectrum using the default-case power spectrum, i.e. taking $P_\text{tt}(k, \mu)$ to be that of the default case. Similarly, we model the +Gzerr+Ifrac$0.05$ power spectrum using the +Gzerr power spectrum. As a result, the ratio is close to unity, indicating that the model catches the effect of interlopers well. For the quadrupole ratio, the spike near $k=0.15\hMpc$ is caused by the zero crossing of the quadrupole.

\begin{figure*}
	\includegraphics[width=1.95\columnwidth]{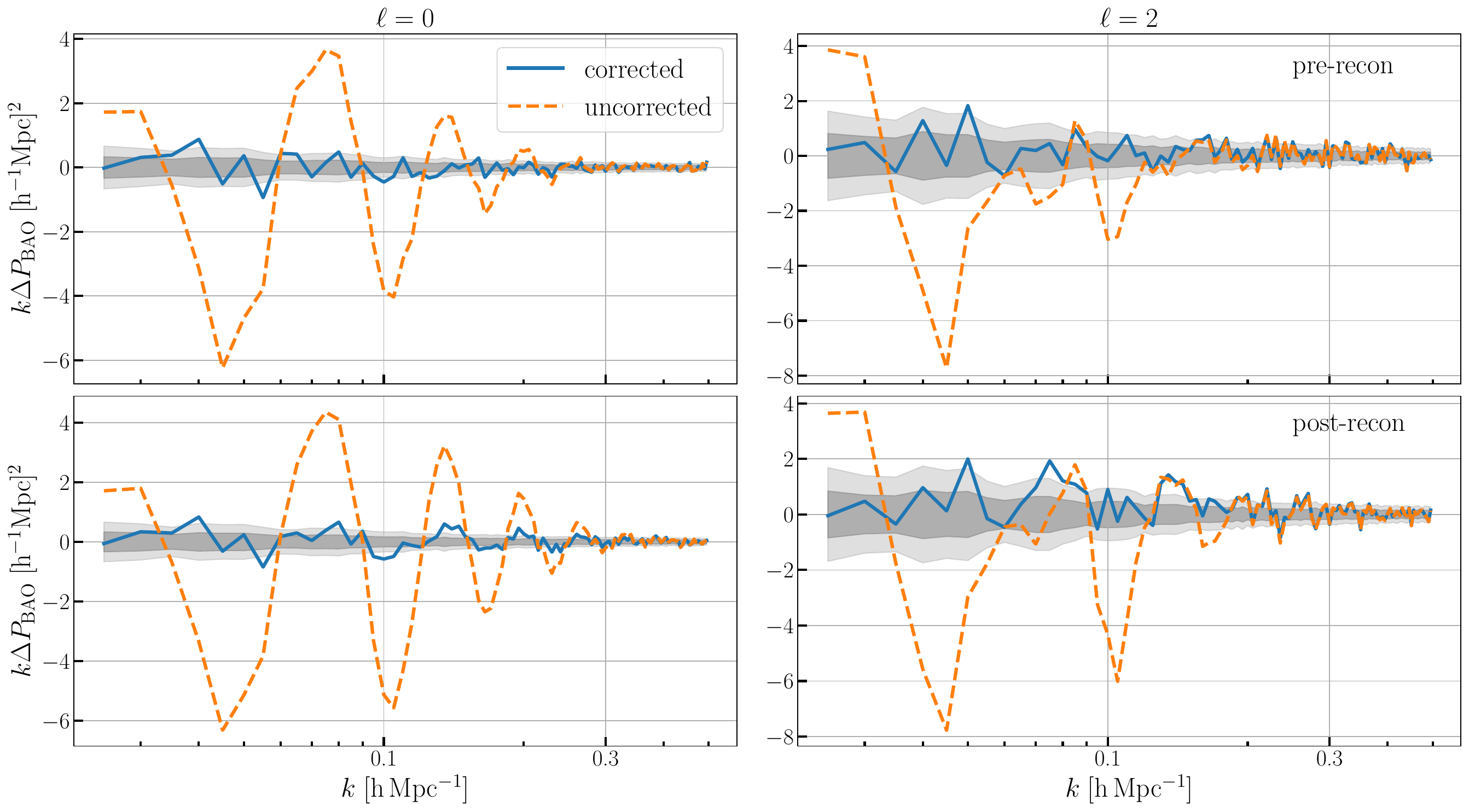}
    \caption{Effect of small-displacement interlopers on the BAO signal. The left and right panels show the monopoles and quadrupoles; the upper and lower panels show the pre- and post-reconstruction cases, respectively. The dashed lines denote the mean difference in the BAO power spectrum multipoles between the default and +Ifrac$0.05$ cases, averaged over 90 realizations, while the solid lines show the case in which we replace the default power spectrum with the model-corrected one the accounts for the interloper effect. After correction, the BAO difference becomes much smaller, comparable to the statistic errors of the mean BAO power spectra in the default case, shown as the gray shaded regions with dark and light shades for $1\sigma$ and $2\sigma$ errors, respectively.}
    \label{fig:Pwnw_int_model_z0.6}
\end{figure*}
Fig.~\ref{fig:Pwnw_int_model_z0.6} shows the impact of small-displacement interlopers on the BAO power spectrum. The shaded regions denote the statistical fluctuations of the mean BAO power spectrum multipoles across 90 realizations for the default case, with $1\sigma$ and $2\sigma$ errors shown in dark and light gray, respectively. Based on equation (\ref{eq:Pbao_int_model}), we model the +Ifrac$0.05$ BAO power spectrum using the default one. We take the difference between the modelled and observed +Ifrac$0.05$ BAO power spectrum multipoles. We average the difference over 90 realizations, and the results are shown as the solid lines. In contrast, the dashed lines show the mean difference between the default and +Ifrac$0.05$ BAO power spectrum multipoles, averaged over all realizations. With the model correction, the BAO difference becomes much smaller and lies mostly within $2\sigma$ region for the pre-reconstruction case shown in the upper panels; a similar behaviour is seen for the post-reconstruction case, shown in the lower panels.       

\begin{figure*}
    \includegraphics[width=1.95\columnwidth]{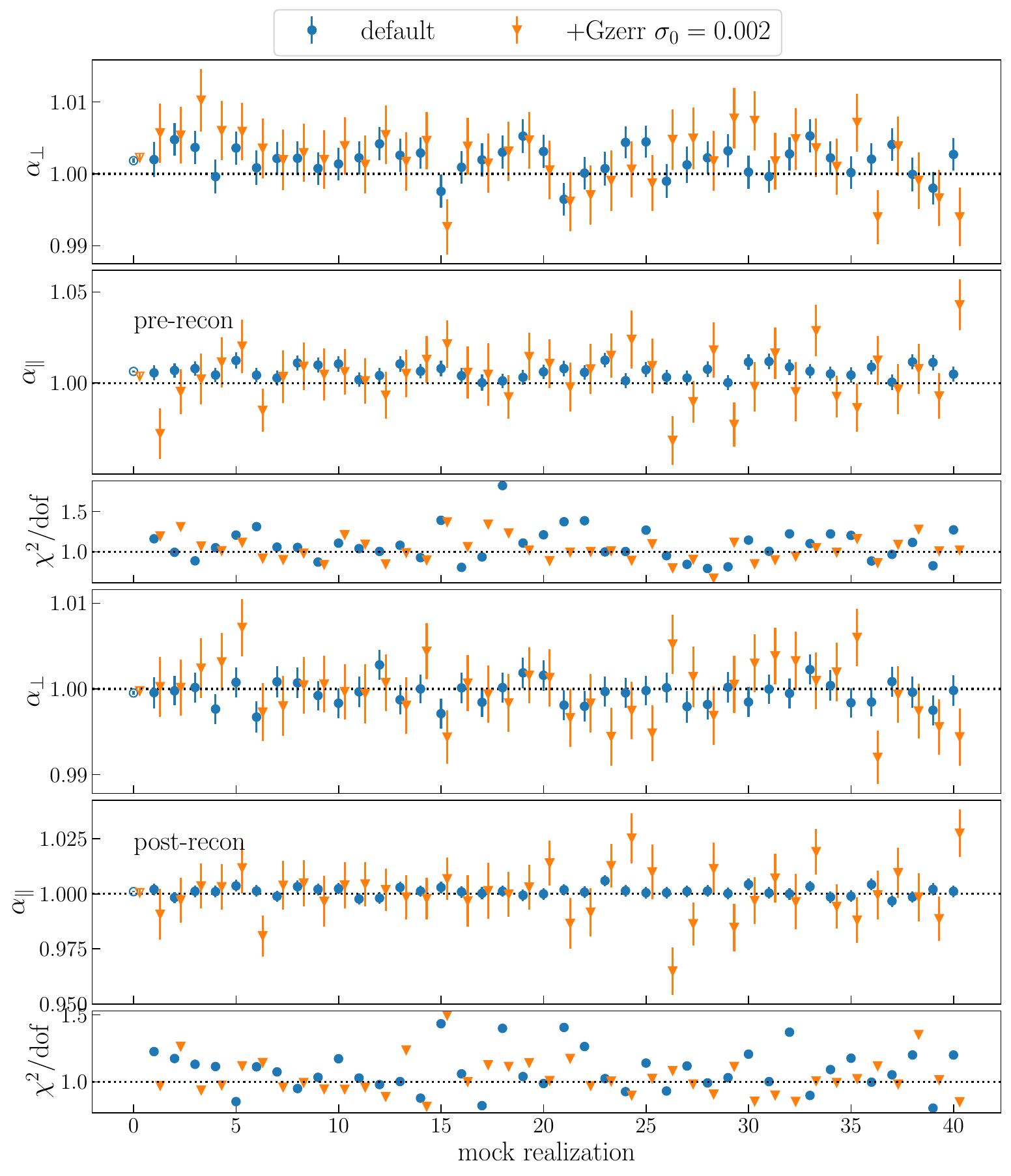}
    \caption{$\alperp$ and $\alpara$ fitted from 40 (out of 90) realizations at $z=1.0$. Circular points correspond to the default case, and triangular points represent the case with Gaussian spec-z uncertainty ($\sigma_0=0.002$). The reduced $\chi^2$ is shown for each fit. Results before and after BAO reconstruction are shown in the upper and lower three panels, respectively. Each solid point is obtained from a fit to the BAO power spectrum monopole and quadrupole of one pair of simulations. We also plot the mean values of $\alpha$ and the standard deviations of the means, shown as the empty points at mock realization 0.}
    \label{fig:alperp_para_z1.0}
\end{figure*}
Fig. (\ref{fig:alperp_para_z1.0}) shows the marginalized means and $1\sigma$ errors of $\alperp$ and $\alpara$ from individual fits. To avoid overcrowding, we show only 40 out of 90 realizations. Circular and triangular points denote the default and +Gzerr cases, respectively. Empty points at mock realization 0 denote the means of the 40 fits. The upper and lower three rows correspond to the cases before and after BAO reconstruction, respectively. For each case, we show the reduced $\chi^2$ of each fit, with $107$ degrees of freedom.

\begin{figure*}
    \includegraphics[width=1.95\columnwidth]{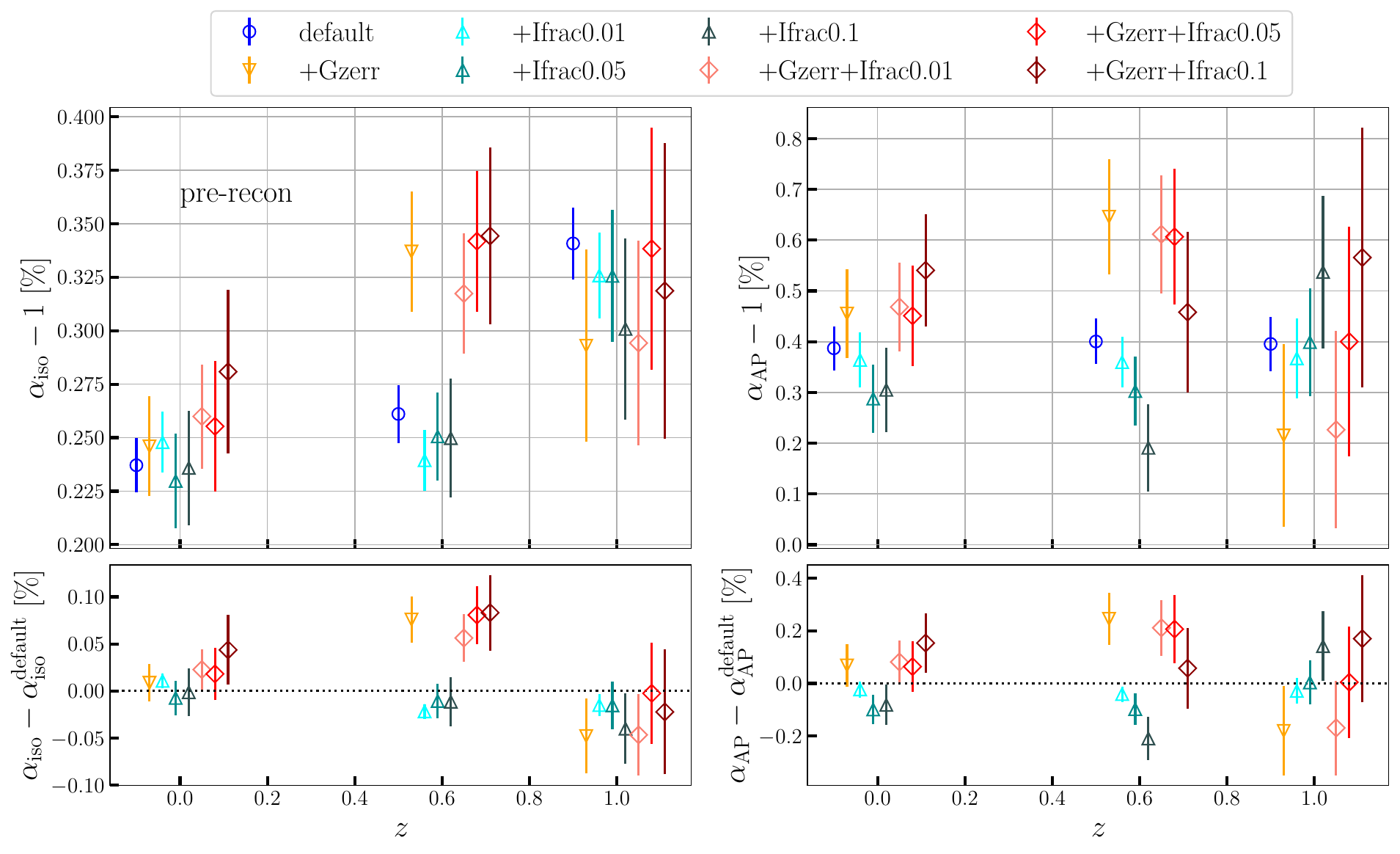}
    \caption{Same as Fig. \ref{fig:alphas_mean_pre} but for $\alpiso$ and $\alpap$ before BAO reconstruction.}\label{fig:alpiso_ap_prerecon}
\end{figure*}

\begin{figure*}
    \includegraphics[width=1.95\columnwidth]{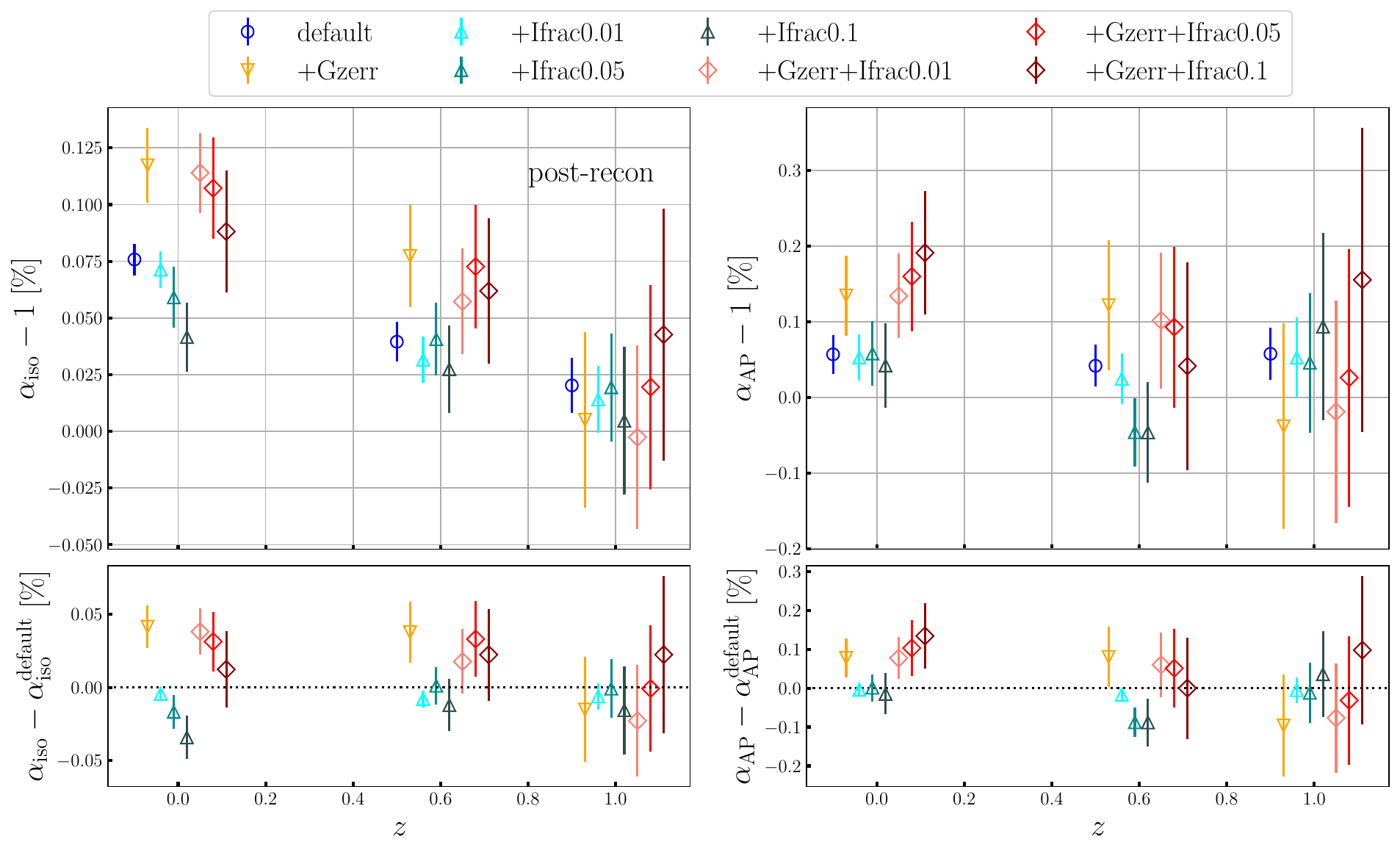}
    \caption{Same as Fig. \ref{fig:alphas_mean_pre} but for $\alpiso$ and $\alpap$ after BAO reconstruction.}\label{fig:alpiso_ap_postrecon}
\end{figure*}
Similar to Figs.~\ref{fig:alphas_mean_pre} and~\ref{fig:alphas_mean_post}, Figs.~\ref{fig:alpiso_ap_prerecon} and~\ref{fig:alpiso_ap_postrecon} show the results of $\alpiso$ and $\alpap$ in the left and right panels, respectively. For each fit, we derive $\alpiso$ and $\alpap$ from the marginalized means of $\alperp$ and $\alpara$ using equation~(\ref{eq:alpiso_ap}). We calculate the means and standard deviations of $\alpiso$ and $\alpap$ across 90 realizations. The systematic biases in $\alpiso$ and $\alpap$ caused by spec-z errors are mostly within $0.1$ per cent and $0.2$ per cent, respectively. 

\begin{figure*}
\includegraphics[width=1.95\columnwidth]{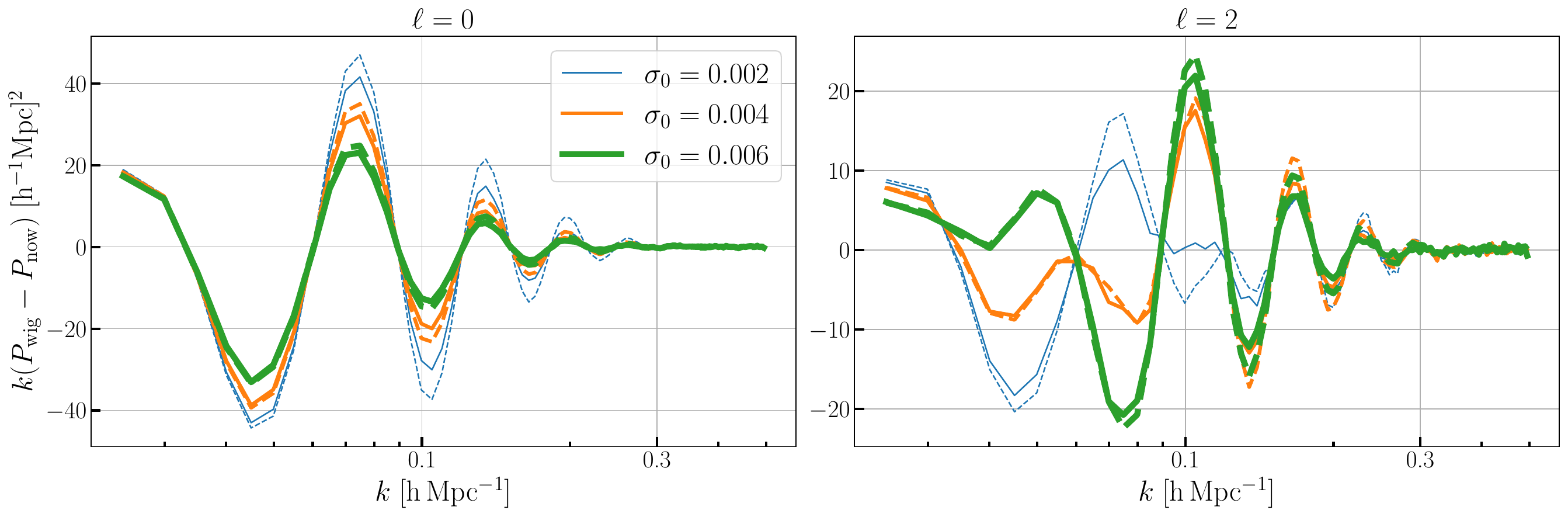}
    \caption{Comparison of BAO power spectrum monopoles (left) and quadrupoles (right) for different Gaussian spec-z uncertainties at $z=0.6$. The line width increases with increasing uncertainty. We plot the mean multipoles averaged over $90$ realizations. The solid and dashed lines correspond to the cases before and after BAO reconstruction, respectively.}\label{fig:bao_z0.6_diffGzerr}
\end{figure*}
Fig.~(\ref{fig:bao_z0.6_diffGzerr}) shows the damping of the BAO power spectrum due to different Gaussian spec-z uncertainties, i.e. $\sigma_0=0.002$, $0.004$, and $0.006$. The left and right panels show the BAO power spectrum monopoles and quadrupoles, respectively. With larger $\sigma_0$, the damping of the BAO signal is stronger. The dashed lines show the results after BAO reconstruction. With larger $\sigma_0$, the difference between pre- and post-reconstruction results is smaller, indicating less improvement in the BAO signal from reconstruction. 

\begin{figure*}
    \centering
    \includegraphics[width=1.995\columnwidth]{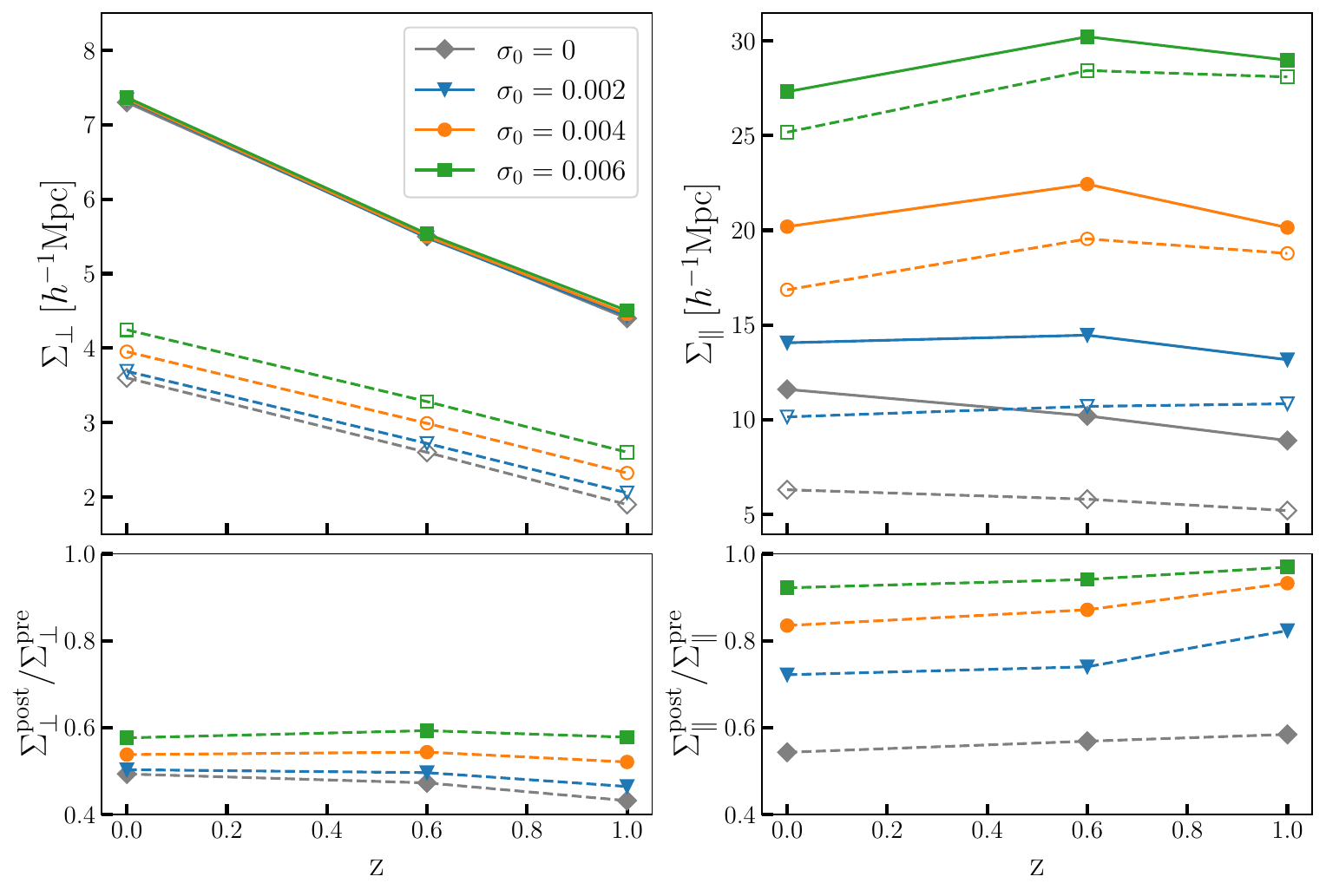}
    \caption{$\Sigmaperp$ and $\Sigmapara$ estimated from the propagators for different Gaussian spec-z uncertainties, shown as different markers. In the upper panels, points connected by solid and dashed lines denote the cases before and after reconstruction, respectively. In the lower panels, we show the ratio of the damping parameters between the post-reconstruction and pre-reconstruction cases. With larger spec-z uncertainty, the ratio is larger, indicating less gain from reconstruction. For comparison, we show the case with RSD only as gray squares.}
    \label{fig:Sigmanl_diffGzerr}
\end{figure*}
Fig.~\ref{fig:Sigmanl_diffGzerr} compares the BAO damping parameters for different spec-z uncertainties. With larger spec-z uncertainty, the propagator from the LoS modes decreases more rapidly as $k$ increases. To avoid the propagator reaching zero, we reduce the $k_p$ value at which $\Sigmapara$ is estimated. We set $k_p=0.2\hMpc$ and $0.12\hMpc$ for the cases with $\sigma_0=0.004$ and $0.006$, respectively. The upper panels show $\Sigmaperp$ (left) and $\Sigmapara$ (right) before and after BAO reconstruction, shown as points connected by solid and dashed lines, respectively. For comparison, we also plot the results from the default case with RSD only as gray points. Before reconstruction, $\Sigmaperp$ is not sensitive to redshift uncertainty, since it is derived from the modes perpendicular to the LoS. While for $\Sigmapara$, related to the modes along the LoS, it increases with increasing redshift uncertainty.

In the lower panels of Fig.~\ref{fig:Sigmanl_diffGzerr}, we show the ratio of the post-reconstruction and pre-reconstruction damping parameters. A larger ratio indicates less gain from reconstruction. For the pessimistic case with $\sigma_0=0.006$, reconstruction reduces $\Sigmaperp$ by $\sim 40$ per cent, but barely reduces $\Sigmapara$. For the optimistic case with $\sigma_0=0.002$, reconstruction reduces $\Sigmapara$ by $20\sim 30$ per cent.

\bsp	
\label{lastpage}

\end{document}